\documentclass{article}

\usepackage{arxiv}
\usepackage{subcaption}
\usepackage[utf8]{inputenc}
\usepackage[T1]{fontenc}
\usepackage{hyperref}
\usepackage{url}
\usepackage{booktabs}
\usepackage{amsfonts}
\usepackage{nicefrac}
\usepackage{microtype}
\usepackage{lipsum}
\usepackage{graphicx}
\usepackage{natbib}
\usepackage{doi}
\usepackage[percent]{overpic}
\usepackage{xcolor}
\usepackage{multirow}
\usepackage{enumitem} 
\usepackage{appendix}
\usepackage{float}

\title{AIFS-DOP: End-to-end medium-range weather prediction from observations alone with machine learning}

\hypersetup{
pdftitle={XXX},
pdfsubject={},
pdfauthor={},
pdfkeywords={},
}

\newbox{\orcid}\sbox{\orcid}{\includegraphics[scale=0.06]{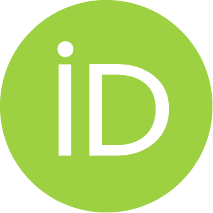}} 

\author{
    {\href{https://orcid.org/0000-0003-1869-3426}{\usebox{\orcid} Ewan Pinnington}}
    \And {\href{https://orcid.org/0000-0002-3662-5382}{\usebox{\orcid} Peter Lean}}
    \And {\href{https://orcid.org/0009-0007-7798-6524}{\usebox{\orcid} Mihai Alexe}}
    \And {\href{https://orcid.org/0000-0002-6070-2544}{\usebox{\orcid} Eulalie Boucher}}
    \And {\href{https://orcid.org/0000-0003-3952-586X}{\usebox{\orcid} Simon Lang}}
    \And {\href{https://orcid.org/0000-0003-2808-0463}{\usebox{\orcid} Patrick Laloyaux}}
    \And {\href{https://orcid.org/0000-0003-3155-480X}{\usebox{\orcid} Gert Mertes}}
    \And {Tomas Kral}
    \AND {\href{https://orcid.org/0000-0002-7374-3820}{\usebox{\orcid} Patricia de Rosnay}}
    \AND {\href{https://orcid.org/0000-0002-1132-0961}{\usebox{\orcid} Matthew Chantry}}
    \And {\href{https://orcid.org/0000-0003-2808-0463}{\usebox{\orcid} Anthony McNally}}
    \And
    European Centre for Medium-Range Weather Forecasts (ECMWF)
}

\begin{document}
\maketitle
\begin{abstract}

We introduce the Artificial Intelligence Forecasting System for Direct Observation Prediction (AIFS-DOP). AIFS-DOP is trained on a 40-year harmonized dataset of gridded observations, without using numerical weather prediction (NWP) reanalysis or model data. The resulting model is competitive with ECMWF's Integrated Forecasting System (IFS) when scored on a one year period of forecasts across 2021/2022. This progress on Direct Observation Prediction represents the first time that a data-driven model, \textit{trained solely on observations}, is competitive with the IFS at medium ranges for several key upper-air and surface headline scores, when verified against observation data.

\end{abstract}

\section{Introduction}
\label{sec:intro}

Data-driven weather forecast models have become increasingly prevalent in recent years. Such models have been predominately trained on ECMWF ERA5 reanalysis data \citep{hersbach2020era5}, produced under the Copernicus programme, and have shown impressive improvements in forecast skill and efficiency (e.g. \cite{keisler2022}, \cite{bi2023nature}, \cite{lam2023}, \cite{lang2024aifs}). ECMWF now runs both the AIFS-Single \citep{lang2024aifs, egusphere-2025-4716} and AIFS-ENS ensemble \citep{lang2026aifs} models operationally, marking a significant moment for machine-learnt weather forecasts. Both operational models were trained under the Anemoi framework (see \cite{lang2024aifs}). Anemoi provides code and tools for the full data-driven forecasting workflow, from the creation of ML-ready datasets to the training of models and running of real-time inference. Anemoi also supplies functionality for cataloguing and tracking of model weights, datasets and training runs to ensure full lineage and reproducibility of the ML workflow, a key requirement for any operational centre.  

The majority of previous data-driven forecasting models, including those mentioned above, are trained on (and initialised from) data assimilation (DA) (re)analysis products which blend observations with physics-based numerical weather prediction (NWP) models using Bayesian methods to find the statistical best estimate to the state of the atmosphere \citep{rabier2000}. These (re)analysis datasets (such as ERA5 \citep{hersbach2020era5}) are extremely powerful for machine learning, providing well curated, high-quality and self-consistent data with many thousands of samples to learn from. By learning to emulate these analysis products, these models are able to minimise errors in medium-range weather forecasts extremely effectively. In contrast, AI Direct Observation Prediction (AI-DOP) aims to learn a skilful weather forecast model from Earth System observations \textit{alone} \citep{mcnally2024}. This offers a route around many of the difficulties with traditional data assimilation, potentially exploiting a wider variety of observations and breaking the dependency of data-driven forecast models on traditional (re)analysis products for training and initialisation. 

Substantial progress has been shown in both the development of observation-driven machine-learned forecasting models and in methods to analyse and understand what they learn.
GraphDOP demonstrated that end-to-end weather forecasting can be learned directly from heterogeneous and sparse satellite and in-situ observations only \citep{alexe2024}.
Observation-driven models develop physically coherent internal representations of Earth System state despite being trained without explicit physical priors \citep{lean2025}. They can capture coupled ocean-atmosphere-cryosphere interactions, such as rapid Arctic sea-ice freezing, heat waves and tropical cyclone-induced cold wakes \citep{boucher2025}. Finally, tools from data assimilation can be adapted to quantify how different observations influence forecast error and make predictions explainable \citep{laloyaux2025}.

Other machine-learned models have shown skillful forecasts achieved through training on a combination of observations and the ERA5 reanalysis dataset \citep{Andrychowicz2023MetNet3,yuval2024,Allen2025Aardvark,ni2025}, thus being able to produce forecasts without relying on inputs from a physics-based NWP model. Until now, a model trained solely on observations (without any reanalysis present) has not displayed day 5 - 10 upper-air/surface scores that are competitive with the benchmark physics-based system, ECMWF's IFS.

Here, we present AIFS-DOP, the first effort to integrate observations into the Anemoi framework, leveraging many of the observation datasets and developments made through the AI-DOP and AIFS research at ECMWF. We show that when training a DOP-type model on gridded Earth System observation data with Anemoi, we can match or surpass the medium-range forecast skill of the ECMWF IFS for several key upper-air and surface headline scores. The AIFS-DOP system uses an increased volume of training data from 10 to 40 years, compared to results previously shown for GraphDOP, achieved through extending backward existing observational datasets by making use of reprocessed satellite Fundamental Data Records (\textit{e.g.,} \citet{eumetsat_hirs_0961}, \citet{eumetsat_ssmt2_0304} and \citet{zou_msu_2013}). The paper is structured as follows; we introduce the dataset and model used in Section~\ref{sec:model and data} before showing results in Section~\ref{sec:results}. Finally, we offer some discussion and outlook for the next developments in AI-DOP and Anemoi in Section~\ref{sec:discussion and conclusion}.

\section{Data and Model}
\label{sec:model and data}

We have curated a dataset containing conventional and satellite observations averaged spatially onto the O96 octahedral reduced Gaussian grid \citep{Wedi2014}, approximately 1-degree/100~km resolution, and archived as an Anemoi dataset, which uses the zarr format \citep{newman_zarr_2024}. The time dimension of the observations is structured as 6-hour slices of data, where if multiple observations are available at a grid cell we select the observation closest to the end of the 6-hour time window; grid-cells containing no observations for a 6-hour time window are assigned a missing value. We have used EUMETSAT and NOAA reprocessed satellite Fundamental Data Records to extend previous GraphDOP datasets further back in time \citep{eumetsat_hirs_0961, eumetsat_ssmt2_0304, zou_msu_2013, knapp_gridsat_2011}, these harmonised datasets share a common period of 1980 to 2022, which form the focus of this study. Details of the observation types and variables contained within this dataset can be seen in Table~\ref{tab:dataset_description}, with instrument acronyms outlined in Table~\ref{tab:instrument_names}. This dataset contains a reduced set of observation types compared to previous GraphDOP results (see \cite{alexe2024, boucher2025, lean2025, laloyaux2025}). We have a similar number of prognostic variables for AIFS-DOP in this study compared to the standard ERA5-trained AIFS ($\sim100$ vs. $\sim90$).  

\begin{figure} 
    \centering
    \includegraphics[width=0.8\textwidth]{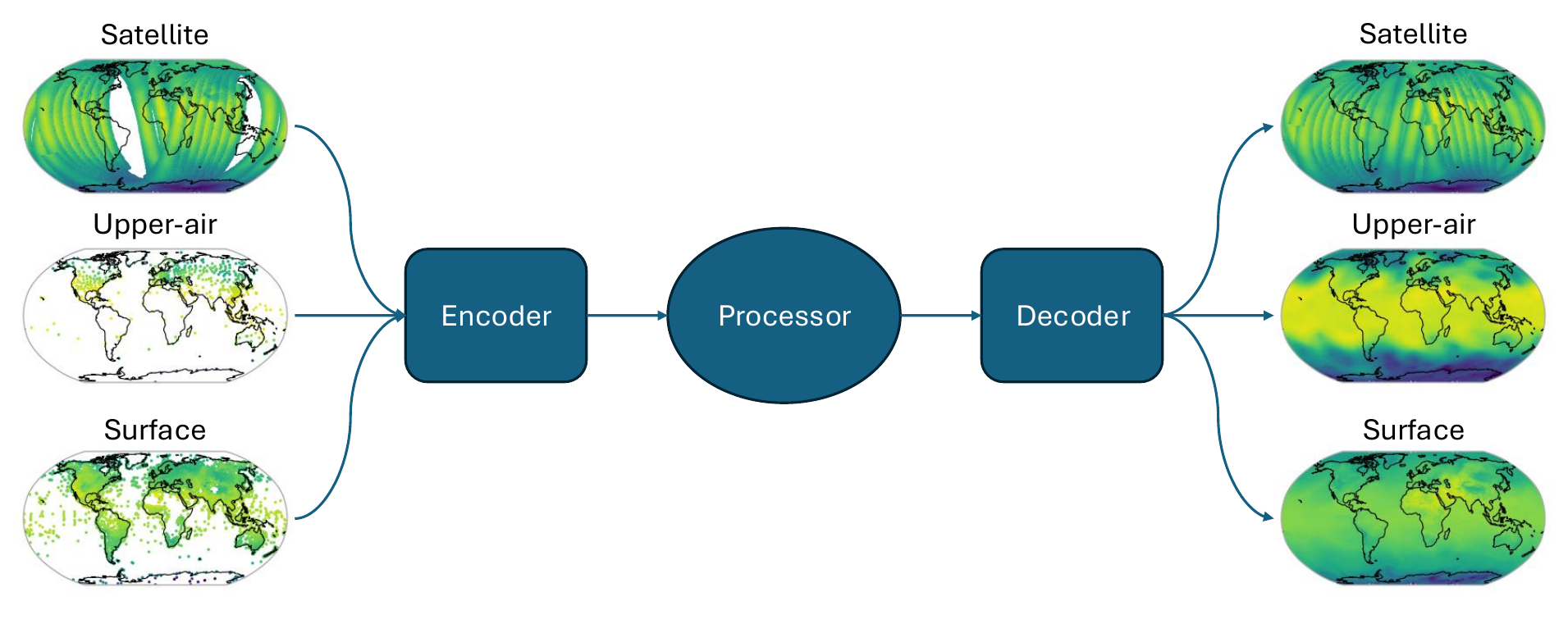}
    \caption{High-level model schematic: A single encoder is used for all observation types. The processor is as described in \citet{lang2024aifs} using a residual connection. Then a single decoder to predict observations out onto a full grid.}
    \label{fig:model}
\end{figure}

\begin{figure} 
    \centering
    \includegraphics[width=0.97\textwidth]{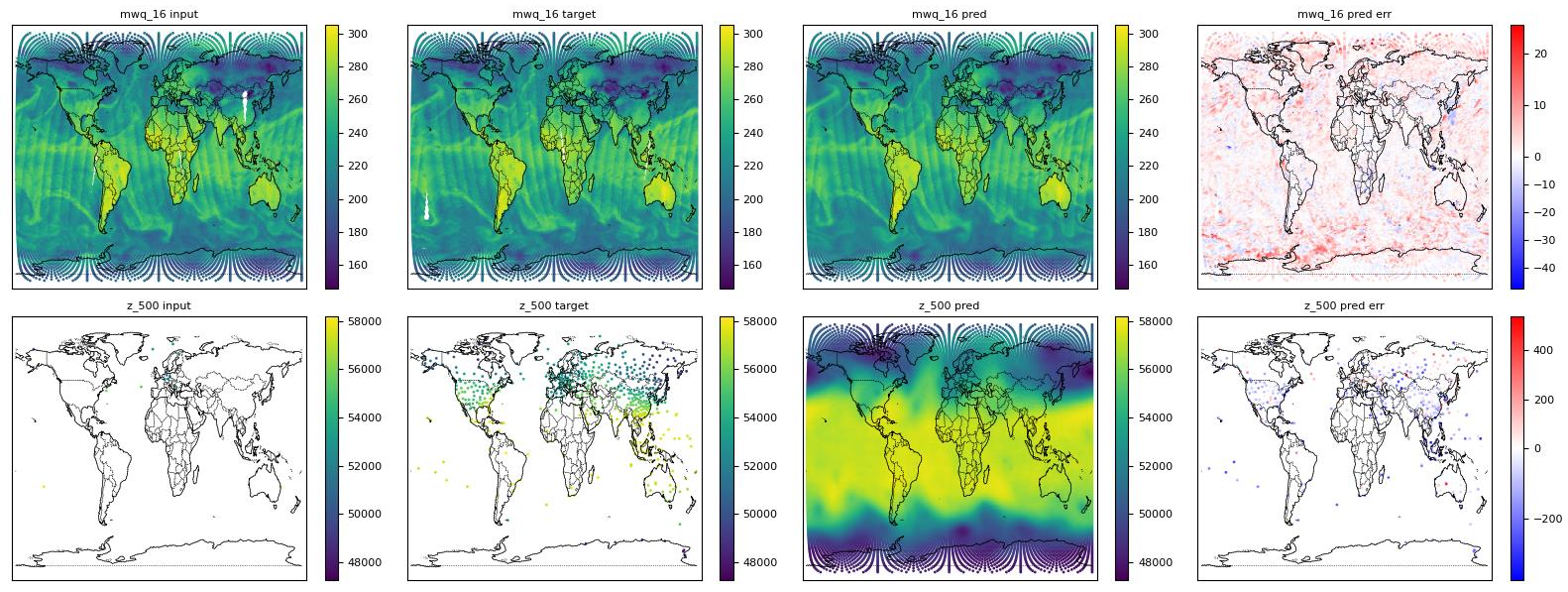}
    \caption{Observation input (left), target (mid-left), prediction (mid-right) and error (right) for ATMS channel 16 (top) and radiosonde geopotential (bottom). A single validation sample is presented here from Anemoi. NaN values in the inputs are replaced by the corresponding mean field values, thus becoming zero after normalization. NaNs in the sparse targets are dropped from the loss before backpropagation, with the model still able to produce full field predictions.}
    \label{fig:train}
\end{figure}

The model uses an encoder-processor-decoder architecture, using a graph-based attention encoder/decoder and a transformer processor with sliding window attention as described in \citet{lang2024aifs}. A schematic of this model is shown in Figure~\ref{fig:model}. This static graph and single encoder/decoder follow the AIFS architecture and are different from the multi-encoder/decoder and dynamic graph architecture of GraphDOP. The model is trained against a mean-squared error loss for years 1981 to 2020 before fine-tuning a 3-day rollout with years 2002 to 2020, with a validation period of January to May 2021. The model takes in two 6-hour gridded observation slices and predicts the next 6-hour slice of observations. We run 4 ``warm-up'' cycles through gridded observation space before forecast initialisation, so that the model sees more windows of ``real'' observations, \textit{e.g.}, for a forecast initialised at 00UTC the model will have access to observations between 00UTC and 18UTC two days earlier (30 hours of observations). In this ``warm-up'' period we are jointly encoding the observations and prior model predictions into the model, in practice this involves blending the tensors of the AIFS-DOP predictions with the observations, retaining observed values wherever available and using the prior prediction elsewhere. We inject some context variables into the decoder before computing the loss (\textit{e.g.}, satellite view angles and locations) and make sure no information on observation locations that will not be available at inference time is shown to the model (\textit{e.g.}, the `future' location of ship or aircraft reports which can contain residual information about the location of atmospheric features such as the jet stream). As the gridded observations contain many missing values, we set these to the corresponding mean field value (thus defaulting to zero after normalization), to allow the network to learn to make meaningful predictions. Figure~\ref{fig:train} shows an example of the inputs, targets and predictions for a validation batch in the first half of 2021. The input values (left) are imputed before entering the model; the missing values in the targets (mid left) are not imputed as we drop these missing values from the loss before back-propagation. The imputation of inputs and dropping of missing values from the loss allow the network to predict full fields (mid right). The right panel shows the error in these full-grid predictions compared to the sparse target observations.

\section{Results}
\label{sec:results}

\begin{figure}
    \centering
    \setlength{\tabcolsep}{0pt}  
    \rotatebox{90}{\makebox[3cm]{ATMS-7}}%
    \begin{subfigure}[b]{0.2363\textwidth}
        \centering
        \includegraphics[width=\textwidth]{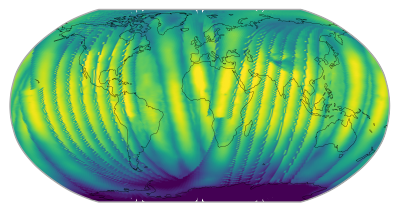}
        \label{fig:w500_1}
    \end{subfigure}%
    \begin{subfigure}[b]{0.2363\textwidth}
        \centering
        \includegraphics[width=\textwidth]{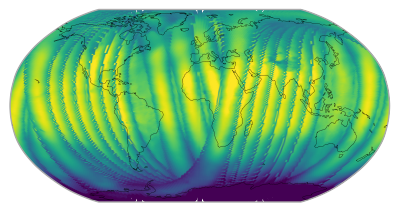}
        \label{fig:w500_5}
    \end{subfigure}%
    \begin{subfigure}[b]{0.2363\textwidth}
        \centering
        \includegraphics[width=\textwidth]{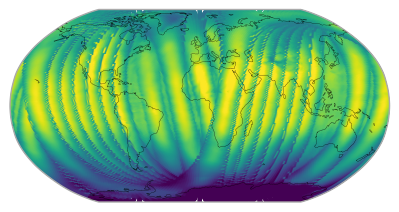}
        \label{fig:w500_10}
    \end{subfigure}%
    \begin{subfigure}[b]{0.2363\textwidth}
        \centering
        \includegraphics[width=\textwidth]{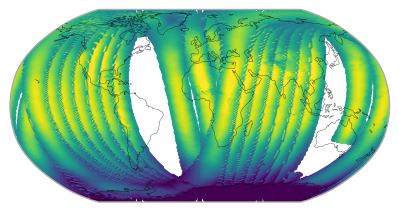}
        \label{fig:w500_obs}
    \end{subfigure}
    
    \vspace{-2.5em}
    
    \rotatebox{90}{\makebox[3cm]{ATMS-16}}%
    \begin{subfigure}[b]{0.2363\textwidth}
        \centering
        \includegraphics[width=\textwidth]{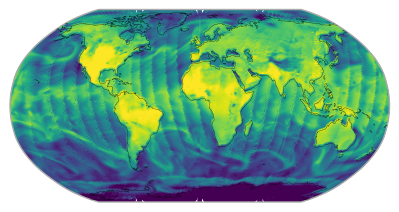}
        \label{fig:w400_1}
    \end{subfigure}%
    \begin{subfigure}[b]{0.2363\textwidth}
        \centering
        \includegraphics[width=\textwidth]{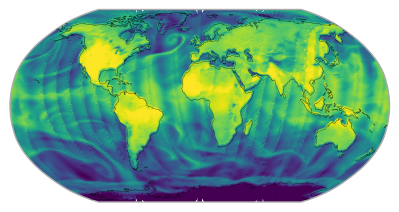}
        \label{fig:w400_5}
    \end{subfigure}%
    \begin{subfigure}[b]{0.2363\textwidth}
        \centering
        \includegraphics[width=\textwidth]{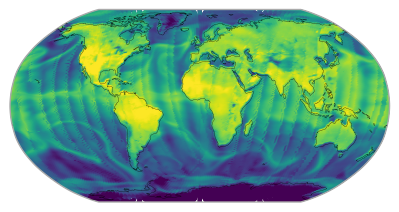}
        \label{fig:w400_10}
    \end{subfigure}%
    \begin{subfigure}[b]{0.2363\textwidth}
        \centering
        \includegraphics[width=\textwidth]{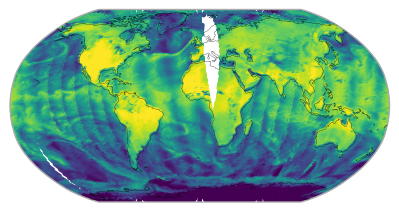}
        \label{fig:w400_obs}
    \end{subfigure}
    
    \vspace{-2.5em}
    
    \rotatebox{90}{\makebox[3cm]{Z-500}}%
    \begin{subfigure}[b]{0.2363\textwidth}
        \centering
        \includegraphics[width=\textwidth]{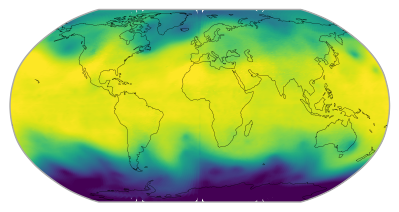}
        \label{fig:z500_1}
    \end{subfigure}%
    \begin{subfigure}[b]{0.2363\textwidth}
        \centering
        \includegraphics[width=\textwidth]{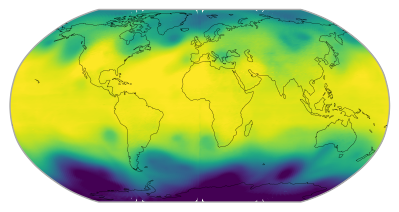}
        \label{fig:z500_5}
    \end{subfigure}%
    \begin{subfigure}[b]{0.2363\textwidth}
        \centering
        \includegraphics[width=\textwidth]{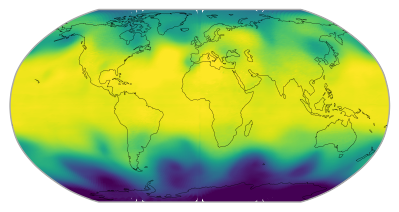}
        \label{fig:z500_10}
    \end{subfigure}%
    \begin{subfigure}[b]{0.2363\textwidth}
        \centering
        \includegraphics[width=\textwidth]{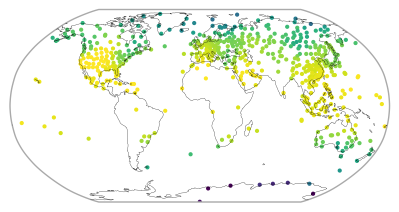}
        \label{fig:z500_obs}
    \end{subfigure}
    
    \vspace{-2.5em}
    
    \rotatebox{90}{\makebox[3cm]{U-250}}%
    \begin{subfigure}[b]{0.2363\textwidth}
        \centering
        \includegraphics[width=\textwidth]{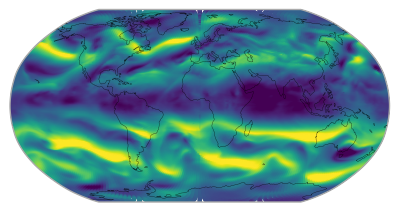}
        \label{fig:u250_1}
    \end{subfigure}%
    \begin{subfigure}[b]{0.2363\textwidth}
        \centering
        \includegraphics[width=\textwidth]{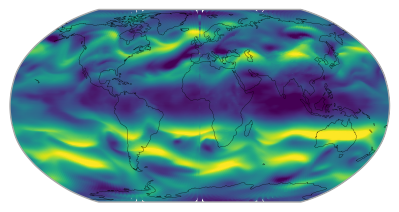}
        \label{fig:u250_5}
    \end{subfigure}%
    \begin{subfigure}[b]{0.2363\textwidth}
        \centering
        \includegraphics[width=\textwidth]{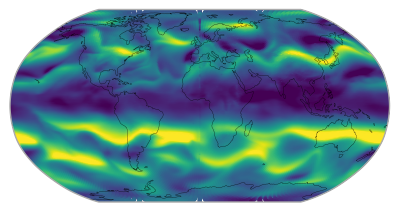}
        \label{fig:u250_10}
    \end{subfigure}%
    \begin{subfigure}[b]{0.2363\textwidth}
        \centering
        \includegraphics[width=\textwidth]{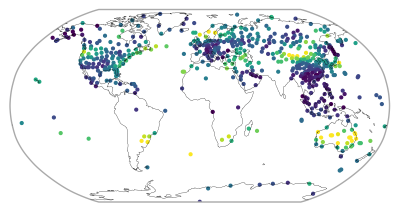}
        \label{fig:u250_obs}
    \end{subfigure}
    
    \vspace{-2.0em}
    
    \rotatebox{90}{\makebox[3cm]{T-850}}%
    \begin{subfigure}[b]{0.2363\textwidth}
        \centering
        \includegraphics[width=\textwidth]{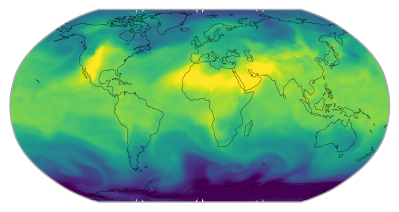}
        \caption*{Day 1}
        \label{fig:t850_1}
    \end{subfigure}%
    \begin{subfigure}[b]{0.2363\textwidth}
        \centering
        \includegraphics[width=\textwidth]{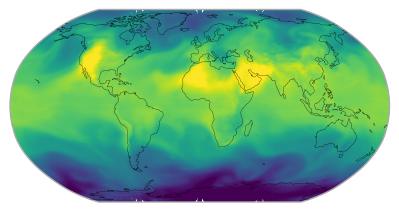}
        \caption*{Day 5}
        \label{fig:t850_5}
    \end{subfigure}%
    \begin{subfigure}[b]{0.2363\textwidth}
        \centering
        \includegraphics[width=\textwidth]{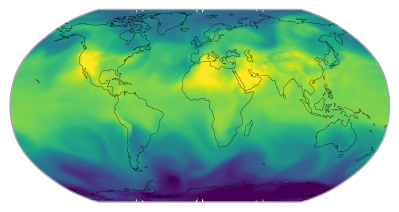}
        \caption*{Day 10}
        \label{fig:t850_10}
    \end{subfigure}%
    \begin{subfigure}[b]{0.2363\textwidth}
        \centering
        \includegraphics[width=\textwidth]{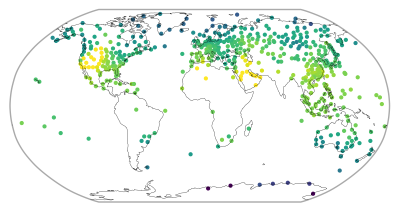}
        \caption*{Observations at Day 10}
        \label{fig:t850_obs}
    \end{subfigure}
    
    \caption{AIFS-DOP predictions at different forecast lead times compared to observations. Forecasts initialised on June 6th, 2021 00~UTC, observations valid at June 16th 00~UTC.}
    \label{fig:model_forecasts}
\end{figure}

In Figure~\ref{fig:model_forecasts}, we show an example forecast from AIFS-DOP for several variables at 24, 120 and 240 hour lead times. We display temperature and humidity sensitive channels from the ATMS instrument (channel 7 and 16 respectively) and three key upper air forecast variables; geopotential at 500~hPa, the u-component of wind at 250~hPa and temperature at 850~hPa. There is consistency across these variables in the location of frontal and large scale weather patterns (\textit{e.g.}, along the jet stream circulation over the North Atlantic, with high U-250 associated to sharp change in Z-500 and T-850 at lower levels), giving us confidence that the model has learnt a robust representation of atmospheric dynamics. The model is still making physically coherent predictions out to day~10 of the forecast with no signs of excessive smoothing. When considering the sparse nature of the upper-air observations used here (see Figure~\ref{fig:train} bottom row), these forecast fields demonstrate the ability of the model to provide complete predictions from sparse data. This is consistent with previous GraphDOP research confirming that such machine-learnt models can develop consistent internal representations of the physical world \citep{lean2025}. The model has also learnt a representation of the limb-effect for the satellite channels (seen in ATMS channel 7 when compared against observations at day 10), finding a regression with the provided viewing zenith angle ``forcing'' variable, in line with what was shown in \citet{lean2025}.

\begin{figure}
    \centering
    \begin{subfigure}[b]{0.33\textwidth}
        \centering
        \includegraphics[width=\textwidth]{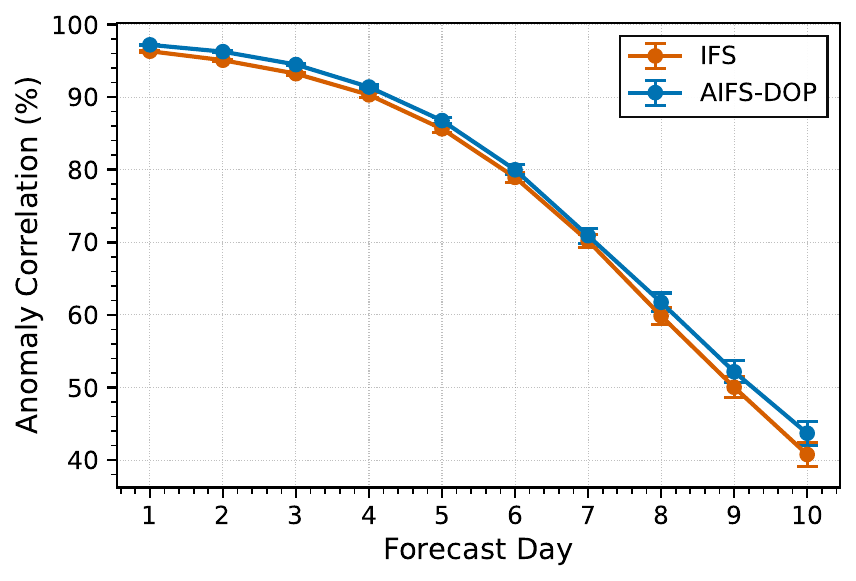}
        \caption{Temperature at 850 hPa}
        \label{fig:t850_nh}
    \end{subfigure}
    \hfill 
    \begin{subfigure}[b]{0.33\textwidth}
        \centering
        \includegraphics[width=\textwidth]{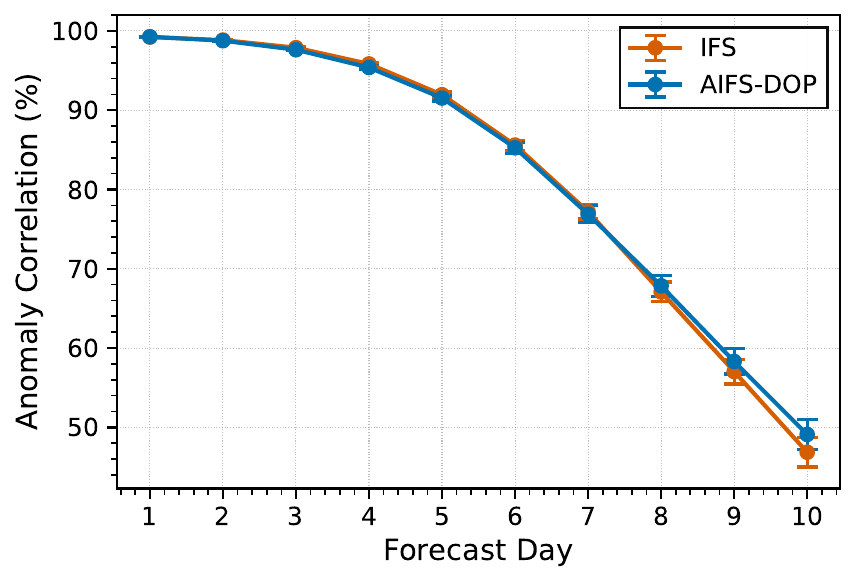}
        \caption{Geopotential at 500 hPa}
        \label{fig:z500_nh}
    \end{subfigure}
    \hfill 
    \begin{subfigure}[b]{0.33\textwidth}
        \centering
        \includegraphics[width=\textwidth]{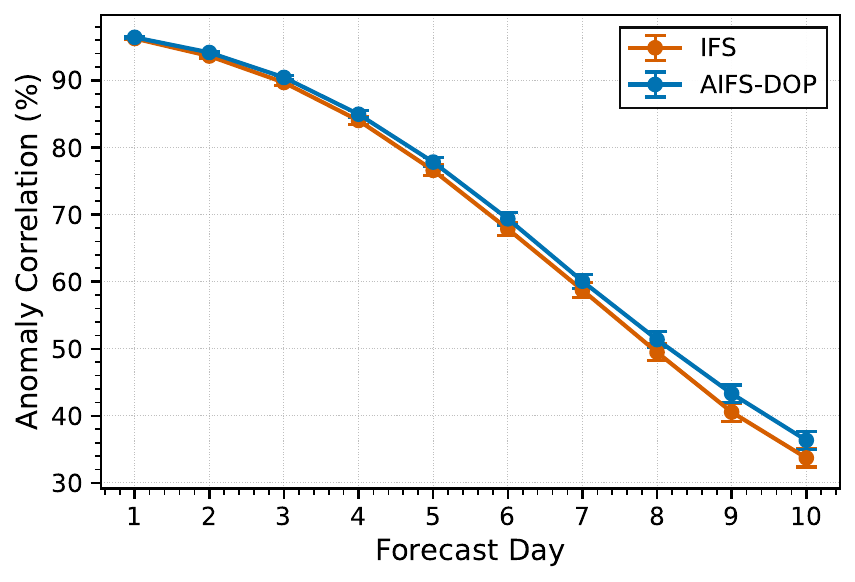}
        \caption{Wind at 250 hPa}
        \label{fig:ff250_nh}
    \end{subfigure}
    \\[1em]
    \begin{center}
        \begin{subfigure}[b]{0.33\textwidth}
            \centering
            \includegraphics[width=\textwidth]{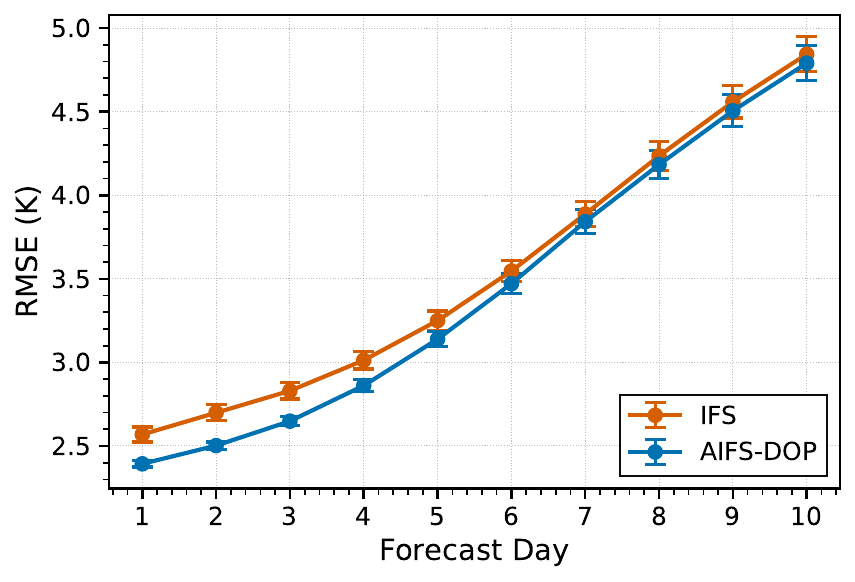}
            \caption{2m Temperature}
            \label{fig:2t0_nh}
        \end{subfigure}
        \hspace{0.05\textwidth}
        \begin{subfigure}[b]{0.33\textwidth}
            \centering
            \includegraphics[width=\textwidth]{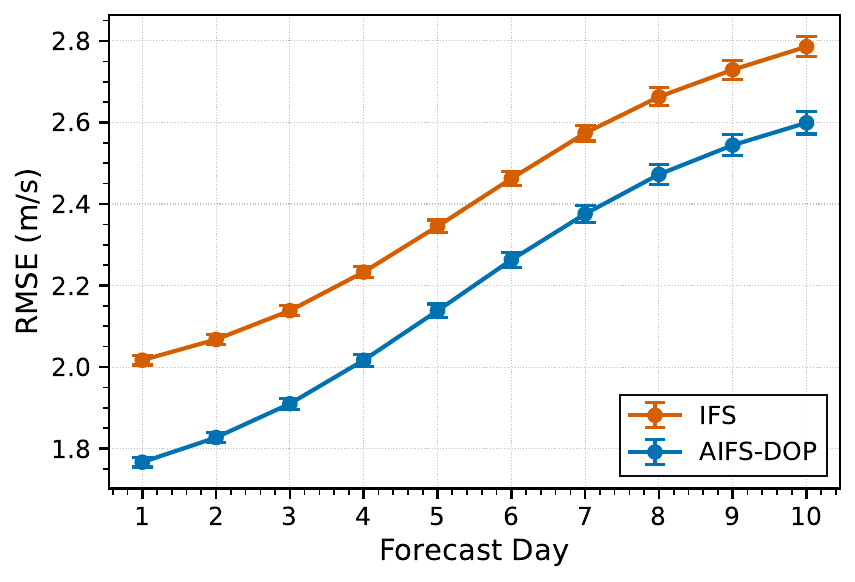}
            \caption{10m Wind}
            \label{fig:10ff0_nh}
        \end{subfigure}
    \end{center}
    \caption{Upper-air anomaly correlation against radiosonde observations (top) and surface root mean square error against SYNOP observations (bottom). Statistics have been averaged for the Northern hemisphere over the period June 2021 to June 2022.}
    \label{fig:nh_scores}
\end{figure}

\begin{figure}
    \centering
    \begin{subfigure}[b]{0.33\textwidth}
        \centering
        \includegraphics[width=\textwidth]{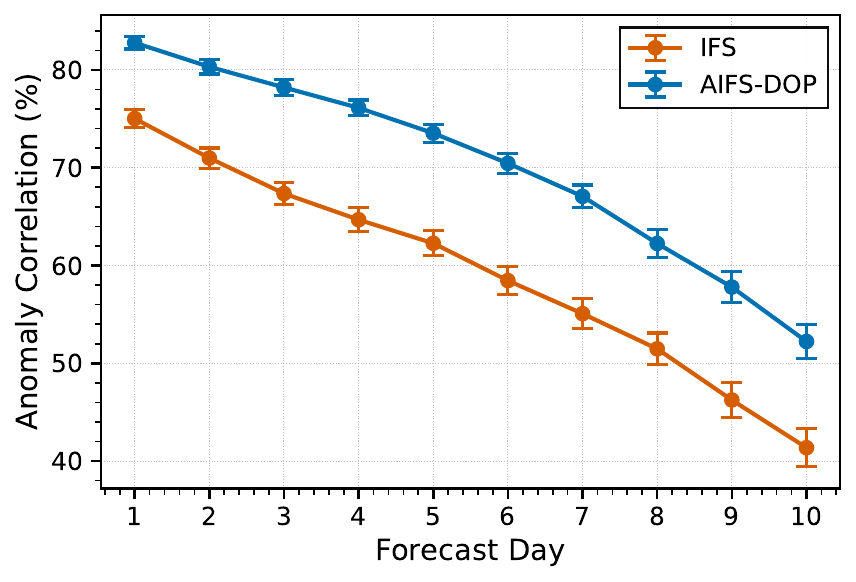}
        \caption{Temperature at 850 hPa}
        \label{fig:t850_trop}
    \end{subfigure}
    \hfill 
    \begin{subfigure}[b]{0.33\textwidth}
        \centering
        \includegraphics[width=\textwidth]{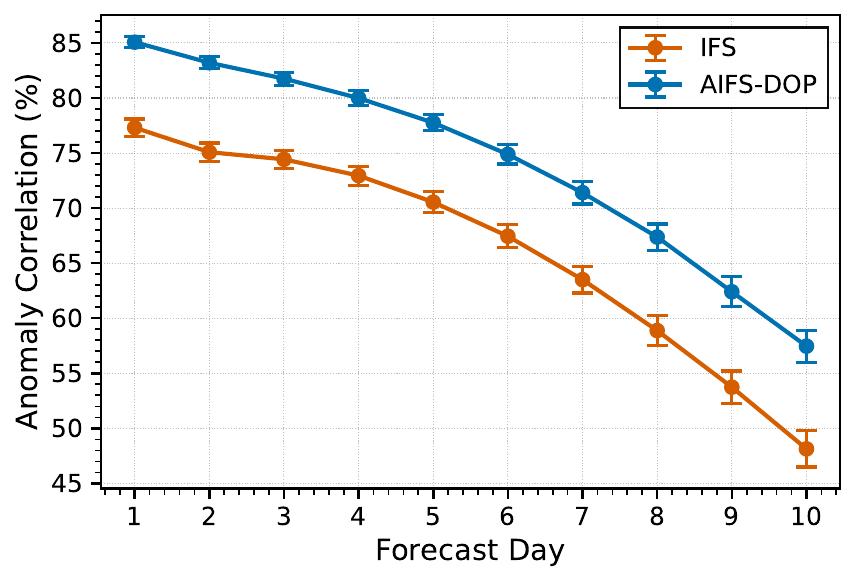}
        \caption{Geopotential at 500 hPa}
        \label{fig:z500_trop}
    \end{subfigure}
    \hfill 
    \begin{subfigure}[b]{0.33\textwidth}
        \centering
        \includegraphics[width=\textwidth]{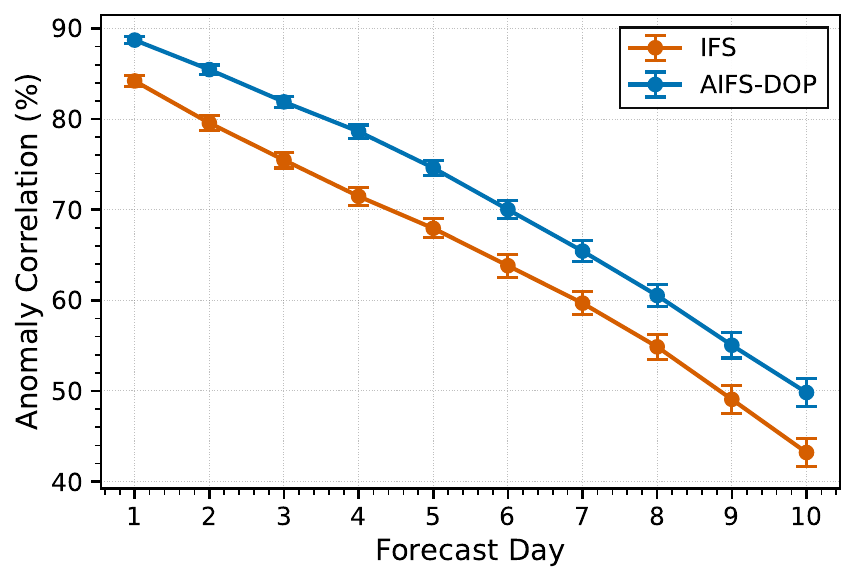}
        \caption{Wind at 250 hPa}
        \label{fig:ff250_trop}
    \end{subfigure}
    \\[1em]
    \begin{center}
        \begin{subfigure}[b]{0.33\textwidth}
            \centering
            \includegraphics[width=\textwidth]{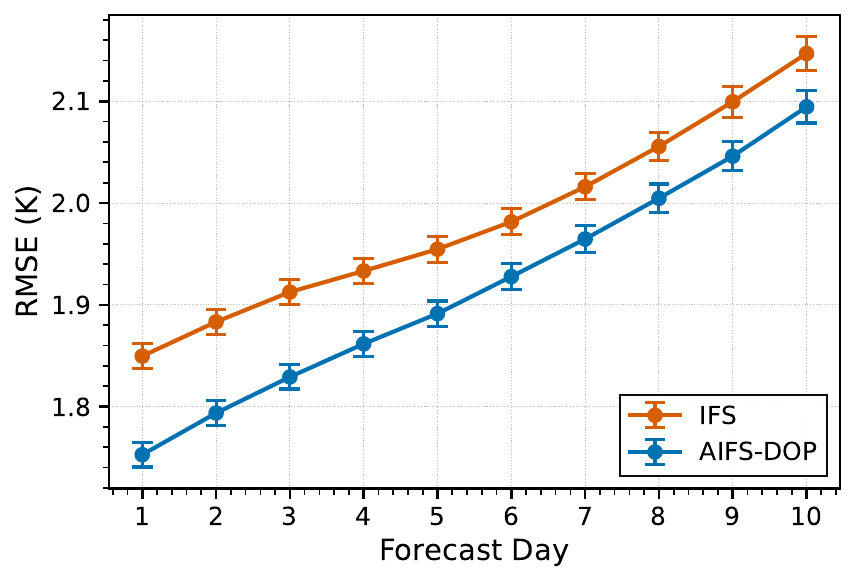}
            \caption{2m Temperature}
            \label{fig:2t0_trop}
        \end{subfigure}
        \hspace{0.05\textwidth}
        \begin{subfigure}[b]{0.33\textwidth}
            \centering
            \includegraphics[width=\textwidth]{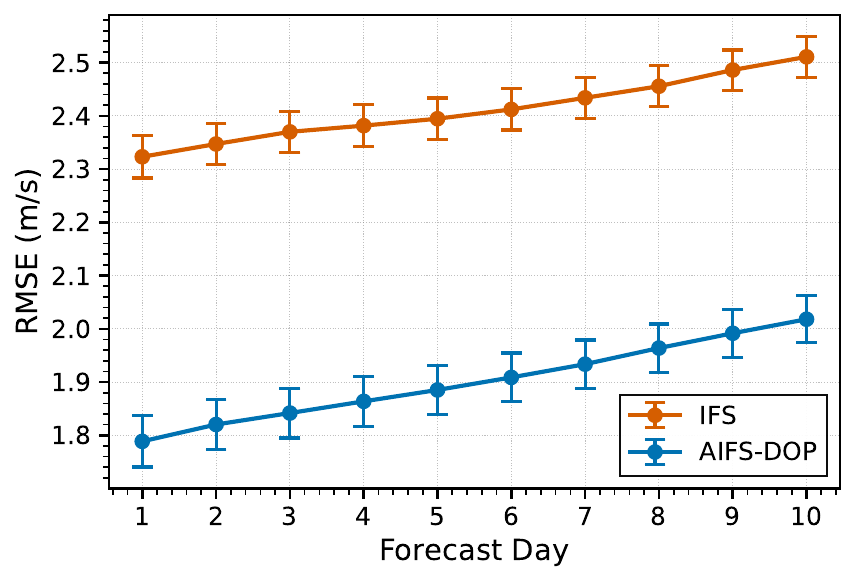}
            \caption{10m Wind}
            \label{fig:10ff0_trop}
        \end{subfigure}
    \end{center}
    \caption{Same as Figure~\ref{fig:nh_scores}, but statistics are computed over the Tropics.}
    \label{fig:trop_scores}
\end{figure}

\begin{figure}
    \centering
    \begin{subfigure}[b]{0.33\textwidth}
        \centering
        \includegraphics[width=\textwidth]{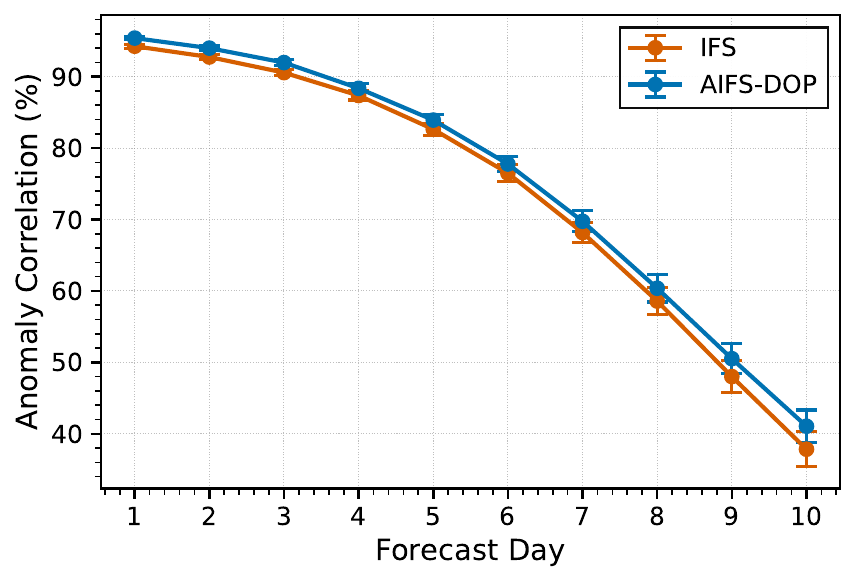}
        \caption{Temperature at 850 hPa}
        \label{fig:t850_sh}
    \end{subfigure}
    \hfill 
    \begin{subfigure}[b]{0.33\textwidth}
        \centering
        \includegraphics[width=\textwidth]{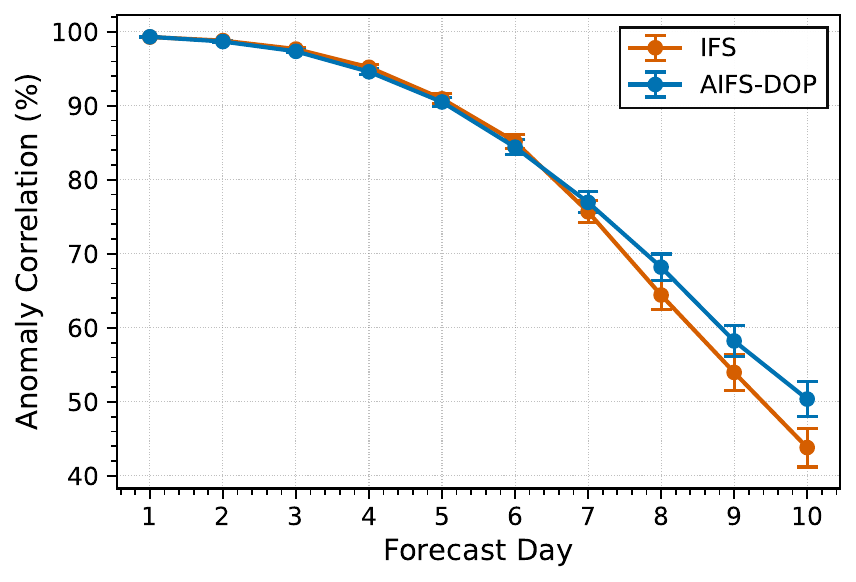}
        \caption{Geopotential at 500 hPa}
        \label{fig:z500_sh}
    \end{subfigure}
    \hfill 
    \begin{subfigure}[b]{0.33\textwidth}
        \centering
        \includegraphics[width=\textwidth]{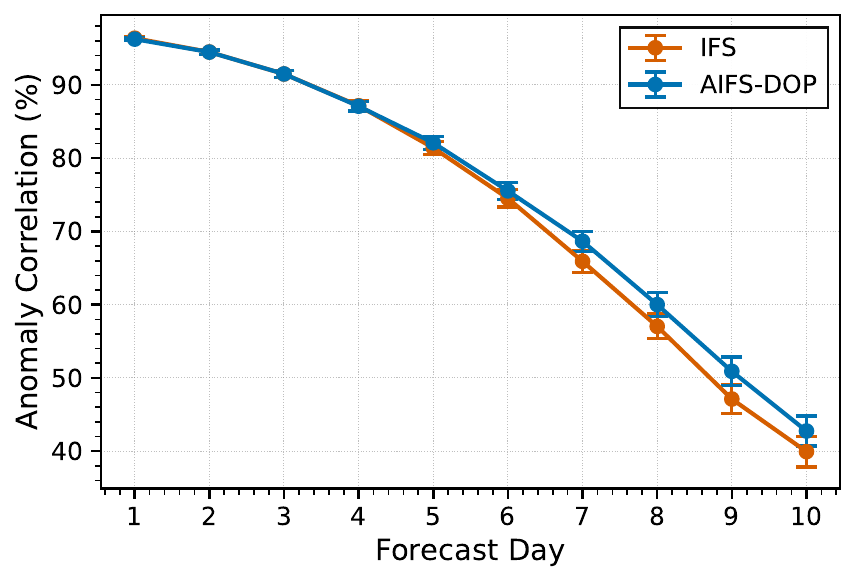}
        \caption{Wind at 250 hPa}
        \label{fig:ff250_sh}
    \end{subfigure}
    \\[1em]
    \begin{center}
        \begin{subfigure}[b]{0.33\textwidth}
            \centering
            \includegraphics[width=\textwidth]{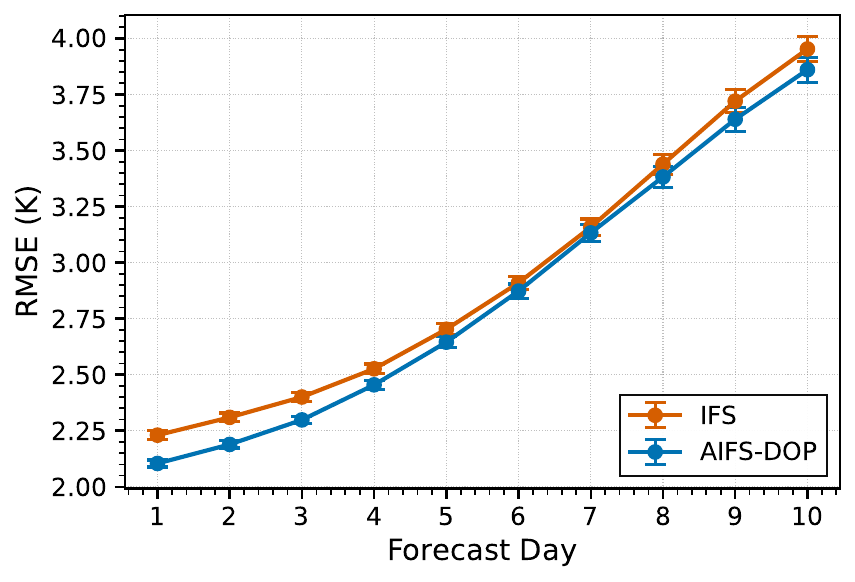}
            \caption{2m Temperature}
            \label{fig:2t0_sh}
        \end{subfigure}
        \hspace{0.05\textwidth}
        \begin{subfigure}[b]{0.33\textwidth}
            \centering
            \includegraphics[width=\textwidth]{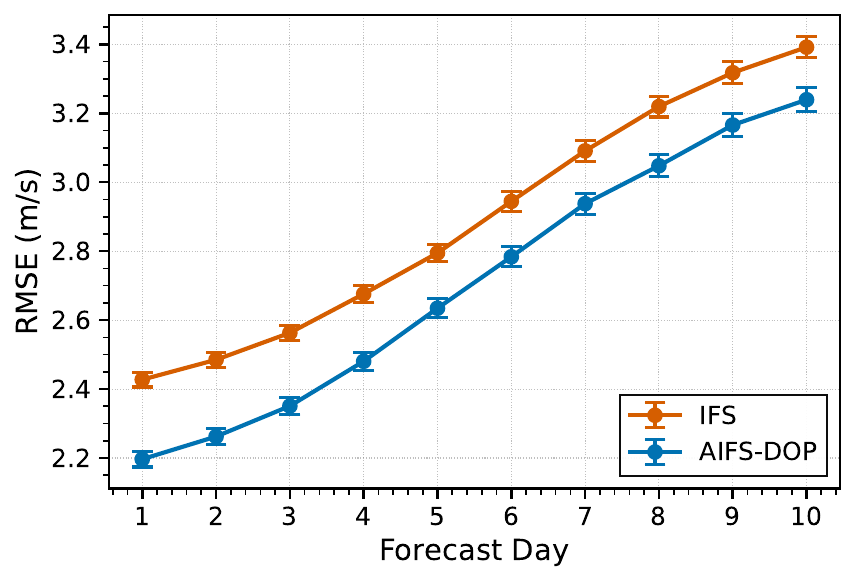}
            \caption{10m Wind}
            \label{fig:10ff0_sh}
        \end{subfigure}
    \end{center}
    \caption{Same as Figure~\ref{fig:nh_scores}, but statistics are computed over the Southern Hemisphere.}
    \label{fig:sh_scores}
\end{figure}

We present anomaly correlation and root-mean squared error skill scores calculated against observations over a full years worth of forecasts (June 2021 to June 2022, using the final year of the harmonised dataset of joint EUMETSAT and NOAA satellite records) initialised at 00 and 12 UTC for the Northern hemisphere, Tropics, and Southern hemisphere in Figures~\ref{fig:nh_scores}, \ref{fig:trop_scores} and \ref{fig:sh_scores}, respectively. We show results for both AIFS-DOP and the corresponding operational IFS forecasts for the period. Considering the Northern hemisphere in Figure~\ref{fig:nh_scores}, we see that in the medium-range AIFS-DOP performance is similar to the IFS with anomaly correlations around 2 to 3~\% higher than the IFS at day~10 for the different variables. For temperature and wind (Figure~\ref{fig:t850_nh} \& \ref{fig:ff250_nh}) AIFS-DOP anomaly correlations are 1~\% higher in the short range, but are very slightly behind the IFS ($\sim 0.5~\%$) for day 5 geopotential (Figure~\ref{fig:z500_nh}). For the surface AIFS-DOP has lower error for the short range and then matches the IFS in the medium range for the 2-meter temperature RMSE (Figure~\ref{fig:2t0_nh}) and lower RMSE at all lead times for 10-meter wind speed (Figure~\ref{fig:10ff0_nh}). For the Tropics in Figure~\ref{fig:trop_scores} we see a higher anomaly correlation coefficient (ACC; 5 to 8\%) and lower RMSE (0.1 to 0.5~K and 0.5 to 0.6 m/s) for AIFS-DOP in both the short and medium ranges for temperature, geopotential, wind, and surface scores. The Tropics signal is dominated by biases that can be minimised effectively when learning from observations. In the Southern Hemisphere (Figure~\ref{fig:sh_scores}) we have a similar pattern as for the Northern Hemisphere but with slight increases in ACC compared to IFS at longer lead times. In Appendix~\ref{appendix:seasonal_scores} we include the scores by season, we can see stronger performance for AIFS-DOP in the Northern/Southern Hemisphere Summer (JJA/DJF) compared to Winter (DJF/JJA), with consistently good performance for the Tropics.


\begin{figure} 
    \centering
    \includegraphics[width=0.98\textwidth]{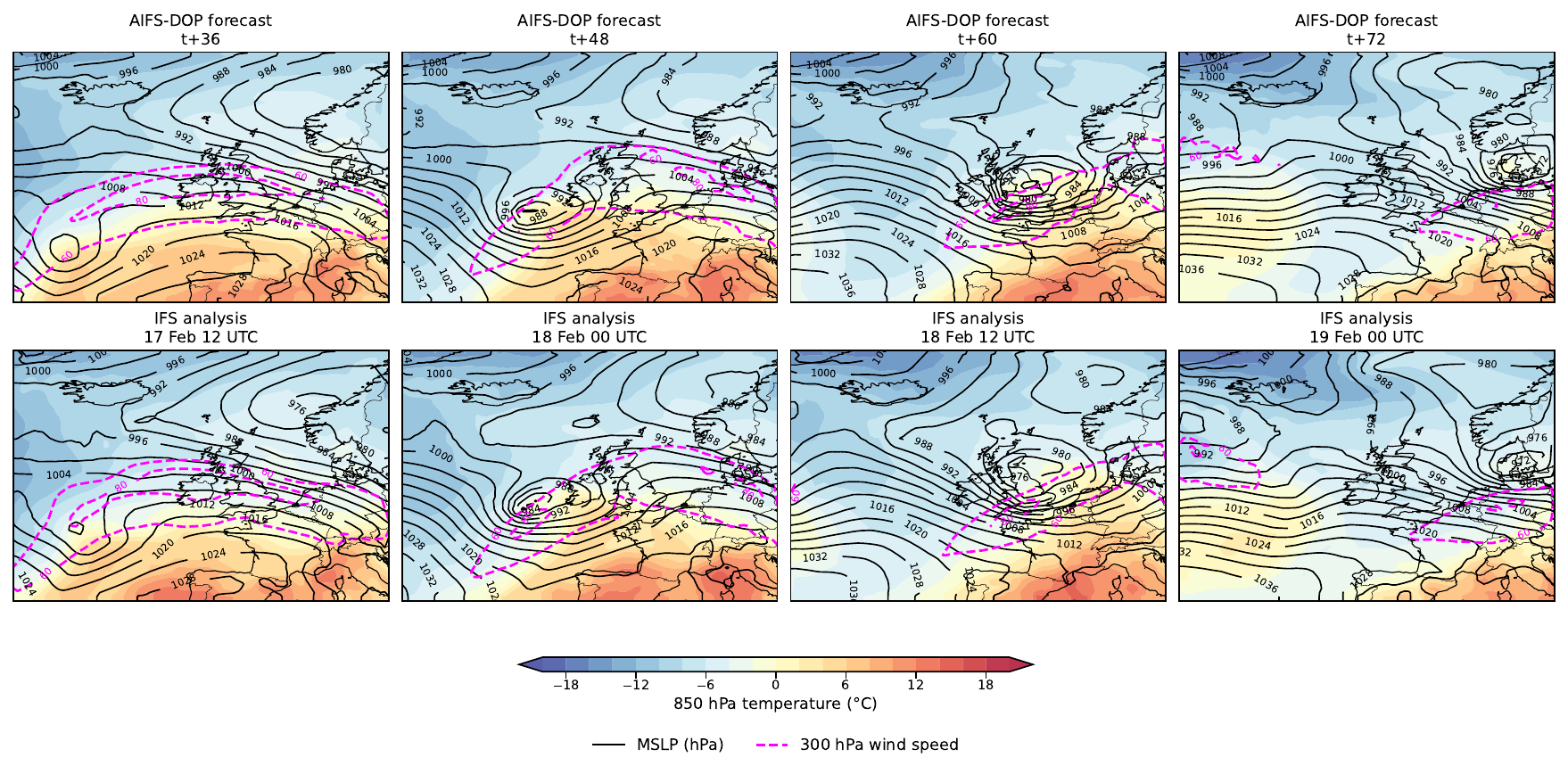}
    \caption{A case study of Storm Eunice for AIFS-DOP. Top row shows AIFS-DOP at lead times of 36, 48, 60 and 72 hours from left to right, bottom row shows corresponding IFS verifying analysis. Pink hatched contours are shown for the 300~hPa wind speed at 60 and 80~m/s.}
    \label{fig:eunice_case}
\end{figure}

In Figure~\ref{fig:eunice_case} we show a case study of Storm Eunice with varying forecast lead times compared to the IFS verifying analysis state. Storm Eunice was a severe weather event over the UK and Ireland in February 2022 that caused 4 fatalities, large scale power-cuts and closure of businesses and schools \citep{volonte_eunice}. Figure~\ref{fig:eunice_case} shows that this event was captured well by the AIFS-DOP forecast, with good agreement displayed against the corresponding IFS analysis state. For a model running at $\sim100$~km horizontal resolution we see well resolved features of temperature, mean sea level pressure and extreme wind speeds out to 72 hours lead time.

\begin{figure} 
    \centering
    \includegraphics[width=0.98\textwidth]{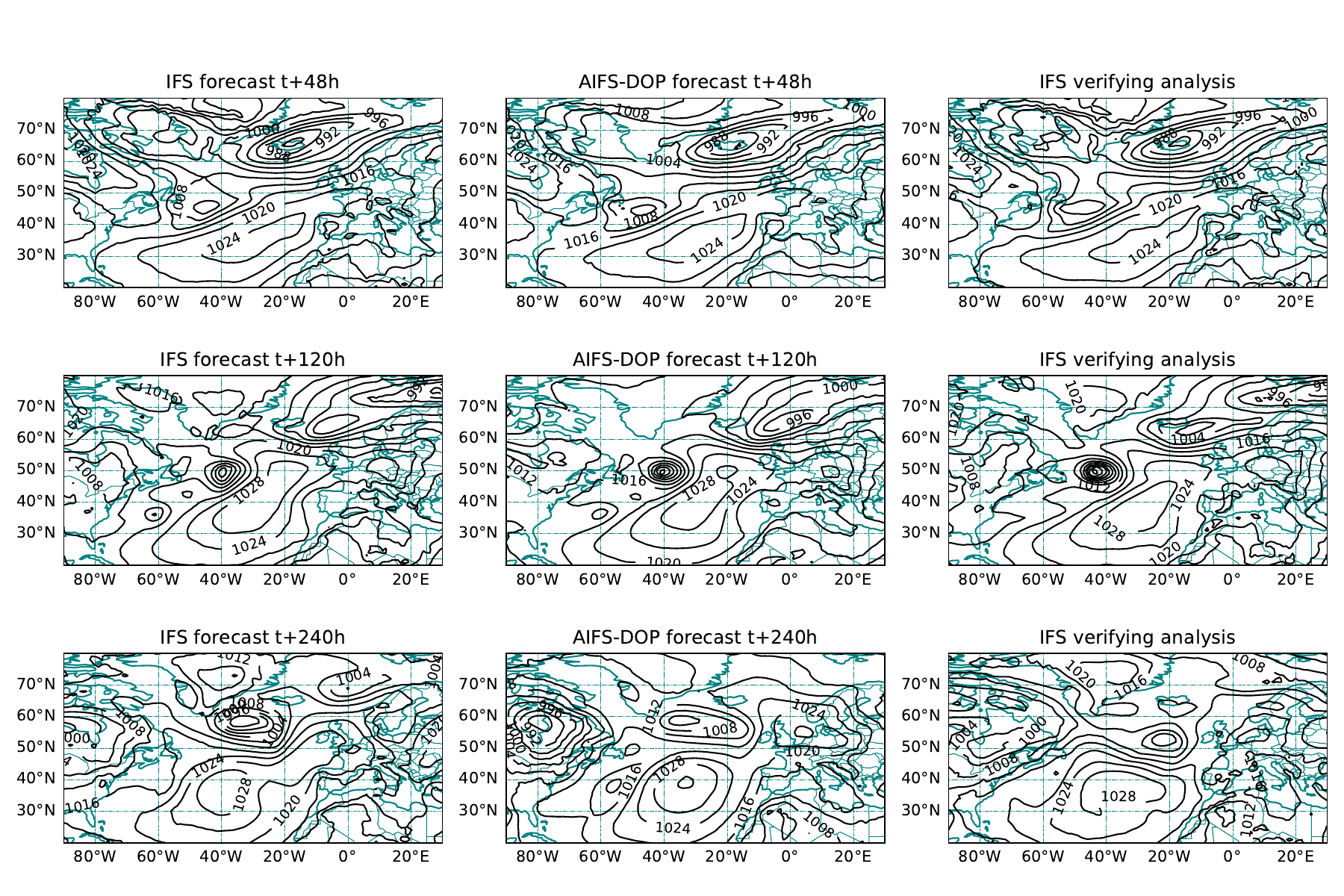}
    \caption{A case study of mean sea level pressure (hPa) over the North Atlantic from the operational IFS forecast (left), AIFS-DOP prediction (centre) and the verifying IFS analysis (right) at lead times of t+48 (top, valid date 11 June 2021 00UTC), t+120 (centre, valid date 14 June 2021 00UTC) and t+240 hours (bottom, valid date 19 June 00UTC). IFS and AIFS-DOP forecasts were both initialised on June 9th, 2021 00 UTC.}
    \label{fig:nhem_msl_case}
\end{figure}

A common feature of data-driven forecast models trained with a mean-squared error loss function is a tendency to smooth forecast fields to minimise the loss at longer lead times. This can lead to a reduction in the smaller scale structure contained within the corresponding forecasts. In order to judge the forecast characteristics of AIFS-DOP we show a case study of mean sea level pressure over the North Atlantic and Southern Ocean in Figure~\ref{fig:nhem_msl_case}. We can see that AIFS-DOP maintains activity at longer lead-times of the forecast with relatively deep areas of low pressure, and captures well the patterns reflected in the IFS verifying analysis. While this issue will require a more systematic evaluation, such case studies already provide some confidence that the model is not suffering from excessive smoothing and is delivering skilful and useful forecasts out to day 10. 

\section{Discussion and Conclusion}
\label{sec:discussion and conclusion}

We have introduced the first data-driven weather forecasting model, trained and initialised solely from observations, that improves on the medium-range skill of the ECMWF IFS for several key upper-air and surface headline scores in the Northern hemisphere, Tropics and Southern Hemisphere when verified against observations. These results demonstrate the progress being made on AI Direct Observation Prediction and through the collaborative Anemoi framework for data-driven model development.

AIFS-DOP shows competitive performance in the medium-range compared to the IFS for the same period. We are comparing the skill of a model trained and run at approximately 100~km horizontal resolution on sparse observations to the IFS at 9~km resolution. We also currently have far fewer observations initialising the AIFS-DOP forecast than the IFS. We give the model five flattened 6-hour windows of observations (one for each observation type in table 1) and directly launch the forecast, so that our model initialised at 00 UTC will have access to observations between 00 UTC and 18 UTC 2 days earlier (00 UTC minus 30 hours). In comparison, the IFS is initialised from the short-window data assimilation analysis at ECMWF which has the advantage of approximately 4 hours of observations beyond the nominal analysis time, so that a forecast initialised at 00 UTC will have access to observations between 21 and 4 UTC the next day. The improvements in AIFS-DOP skill compared to GraphDOP are mainly attributed to the increase of training data volume from $\sim10$ to $\sim40$ years, made possible by the use of reprocessed satellite fundamental data records and the reductions in training time with the simplification of observation gridding. It is possible that the flattening of the time dimension onto 6-hour slices of gridded data and forcing the model to predict out onto a full grid, then autoregressively cycling these gridded predictions, also contribute to help to stabilise rollout. Once trained, the AIFS-DOP model takes less than one minute to make a 10-day forecast, including reading input data and writing output data. As AIFS-DOP can launch instantaneously from the latest observations (without having to wait for the latest data assimilation analysis cycle to finish), this means that we could potentially produce forecasts with virtually zero latency (from observations being available) at arbitrary initialisation times, and provide much quicker access to weather forecasts and associated warnings.

In this study, we focus on a reduced set of observation types while exploring the impact of extending the training period backward in time. We believe that there is significant potential to improve scores further by increasing the amount of training/initialisation data through the use of additional satellite missions and sensing technologies (\textit{e.g.,} radio-occultation, visible, microwave imagers, hyper-spectral infrared, etc.) and dedicated conventional observing networks in data-sparse regions (\textit{e.g.,} \citet{essd-15-411-2023}). Based on previous experience with the AIFS, we expect that an increase in the horizontal resolution of the gridded input/output observations will improve forecast skill at shorter lead times and potentially for surface variables. In future applications, we will investigate training AIFS-DOP against a probabilistic loss, instead of the mean-squared error version presented here, to better understand the potential of ensemble forecasting directly from observations \citep{lang2026aifs,Price2025GenCast}. Probabilistic training will also alleviate the impact of smoothing on forecast verification.

Further promising research directions include increasing temporal resolution, and understanding which observation types benefit from adopting dynamic (time-varying) graphs (as implemented in GraphDOP \citep{alexe2024}) to support arbitrary observation locations for both inputs and outputs once they are available within the Anemoi ecosystem. Assessing the forecast skill for other Earth System components, following \cite{boucher2025}, will be important. In addition, exploring the use of such systems to produce climate reanalysis datasets from observations \textit{alone} will be informative to understand the limits with varying observational networks over time. We expect that the growing Anemoi ecosystem will facilitate the integration of new data sources, model architectures, and training strategies, accelerating progress in observation-driven Earth System prediction.

The results presented here represent a significant milestone for AI-DOP research, showing that it is possible for a data-driven model trained on observations alone to show skill competitive to that of the ECMWF IFS, without any dependency on reanalysis or operational analysis products from traditional NWP centres. Further work will be required to operationalise such machine-learnt observation-driven systems. Substantial effort was required to process and harmonise the datasets (largely based on reprocessed satellite observations from EUMETSAT and NOAA) used for training and inference of the model presented in this study. We will need to replicate these efforts to produce consistent datasets up to real-time appropriate for launching operational forecasts. ECMWF already acquires and processes large volumes of observational data in real time for its world-leading data assimilation system, and leveraging this existing infrastructure for machine-learnt observation-driven forecast models presents a promising new direction for an operational centre.

\section*{Code and Data availability}
AIFS-DOP was trained using the open-source Anemoi framework \url{https://github.com/ecmwf/anemoi}. Much of the data documented here is available from EUMETSAT and NOAA, see Table~\ref{tab:dataset_description}. 

\section*{Acknowledgements}
We extend thanks to Viju John and Roope Tervo at EUMETSAT for help accessing and discussions on the fundamental data records. Ewan Pinnington’s contribution is funded by the CERISE project (grant agreement No101082139). CERISE is funded by the European Union. Views and opinions expressed are however those of the author(s) only and do not necessarily reflect those of the European Union or the Commission. Neither the European Union nor the granting authority can be held responsible for them. We acknowledge PRACE for awarding us access to Leonardo, CINECA, Italy. We acknowledge the EuroHPC Joint Undertaking for awarding this work access to the EuroHPC supercomputer JUPITER, hosted by Jülich Supercomputing Centre (JSC).

\bibliographystyle{plainnat}
\bibliography{references}
\appendix

\section*{Appendix}

\section{Specification of training datasets}

Table \ref{tab:dataset_description} lists the datasets that were used to train the model described in Section \ref{sec:model and data}.

\begin{table}[htbp!]
\caption{Description of curated observation dataset}
\label{tab:dataset_description}
\centering
\begin{tabular}{|p{2.5cm}|p{2.5cm}|p{1.8cm}|p{2.5cm}|p{4.5cm}|}
\hline
\textbf{Category} & \textbf{Instruments} & \textbf{Period} & \textbf{Variables} & \textbf{Details} \\
\hline
Infrared Sounder & HIRS & 1980--2021 & Brightness Temperatures & EUMETSAT Fundamental Data Record \citep{eumetsat_hirs_0961} \\
\hline
Microwave Sounders & MSU \newline SSM/T-2 \newline AMSU-A \newline AMSU-B \newline MHS \newline ATMS & 1980--2005 \newline 1994--2005 \newline 1998--2021 \newline 1998--2014 \newline 2005--2021 \newline 2012--2021 & Brightness Temperatures & MSU taken from NOAA Climate Data Record \citep{zou_msu_2013}, SSM/T2 taken from EUMETSAT Fundamental Data Record \citep{eumetsat_ssmt2_0304} \\
\hline
Surface Observations & SYNOP, Buoys, Ships & 1980--2021 & 2t, 2d, msl, 10u, 10v, sst & Existing ECMWF data archive \\
\hline
Upper-air Observations & Radiosonde, Aircraft, AMV & 1980--2021 & t, u, v, z, q on pressure levels & Existing ECMWF data archive\\
\hline
Geostationary Satellite & GridSat & 1980--2021 & Brightness Temperatures & NOAA Climate Data Record \citep{knapp_gridsat_2011} \\
\hline
\end{tabular}
\end{table}

\section{Instrument acronyms}
\label{appendix:acronyms}

Table \ref{tab:instrument_names} lists the full names of the satellite instruments that were used in the present study.

\begin{table}[ht]
\centering
\caption{Instrument name definitions.}
\label{tab:instrument_names}
\begin{tabular}{ll}
\toprule 
Acronym & Full name \\
\midrule
MSU & Microwave Sounding Unit \\
AMSU-A & Advanced Microwave Sounding Unit-A \\
ATMS   & Advanced Technology Microwave Sounder \\
SSM/T-2 & Special Sensor Microwave Humidity-2 \\
AMSU-B & Advanced Microwave Sounding Unit-B \\
MHS    & Microwave Humidity Sounder \\ 
HIRS & High Resolution Infrared Radiation Sounder \\
AMV & Atmospheric Motion Vectors \\
\bottomrule
\end{tabular}
\end{table}

\section{Seasonal Scores}
\label{appendix:seasonal_scores}

\begin{figure}
    \centering
    \begin{subfigure}[b]{0.18\textwidth}
        \centering
        \includegraphics[width=\textwidth]{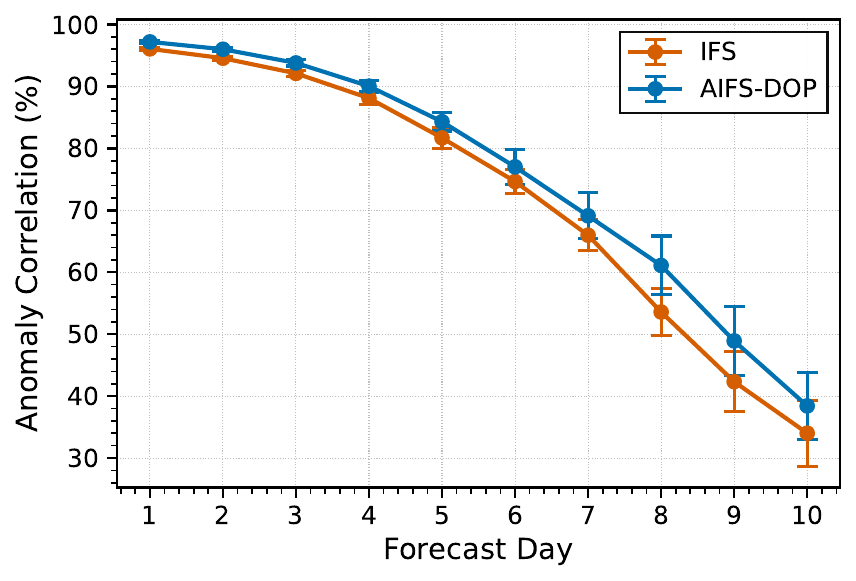}
        \caption{NH T-850}
        \label{fig:t850_nh_jja}
    \end{subfigure}
    \hfill 
    \begin{subfigure}[b]{0.18\textwidth}
        \centering
        \includegraphics[width=\textwidth]{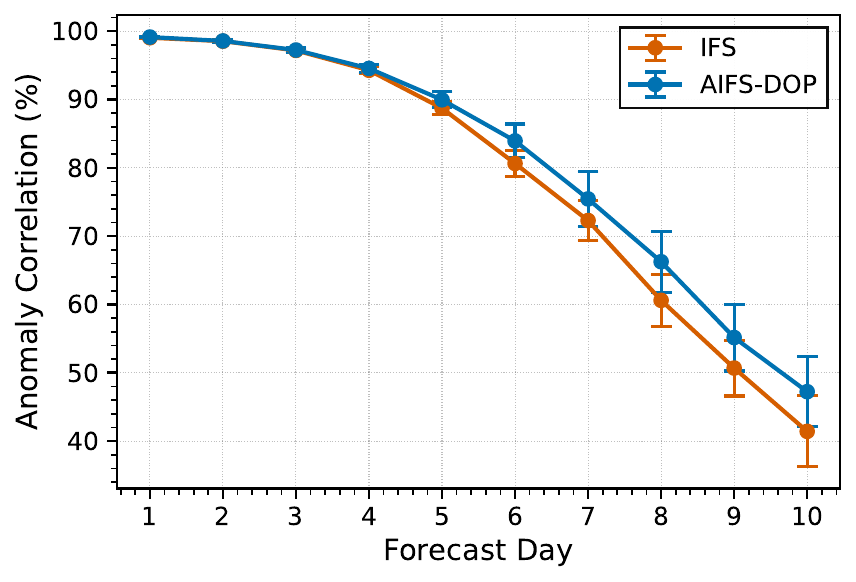}
        \caption{NH Z-500}
        \label{fig:z500_nh_jja}
    \end{subfigure}
    \hfill 
    \begin{subfigure}[b]{0.18\textwidth}
        \centering
        \includegraphics[width=\textwidth]{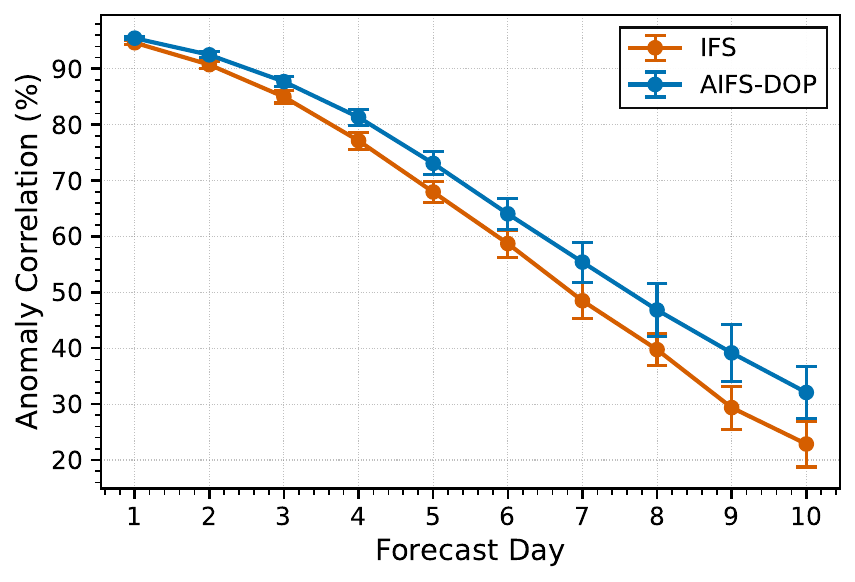}
        \caption{NH Wind-250}
        \label{fig:ff250_nh_jja}
    \end{subfigure}
    \hfill
    \begin{subfigure}[b]{0.18\textwidth}
        \centering
        \includegraphics[width=\textwidth]{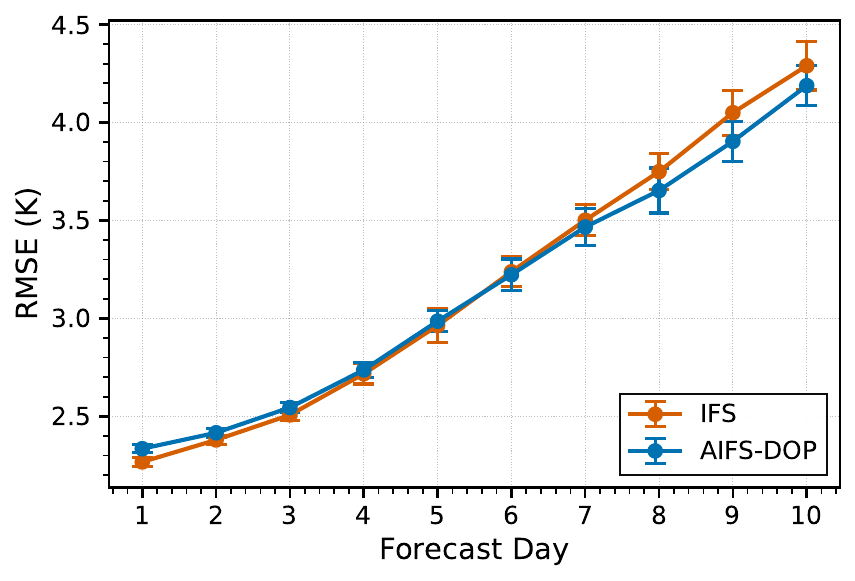}
        \caption{NH T2m}
        \label{fig:2t0_nh_jja}
    \end{subfigure}
    \hfill
    \begin{subfigure}[b]{0.18\textwidth}
        \centering
        \includegraphics[width=\textwidth]{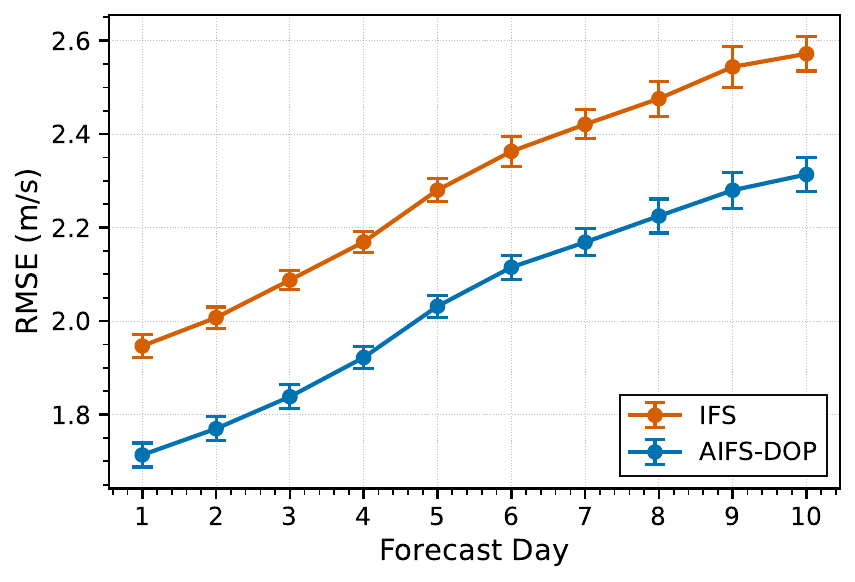}
        \caption{NH 10m Wind}
        \label{fig:10ff0_nh_jja}
    \end{subfigure}

    \begin{subfigure}[b]{0.18\textwidth}
        \centering
        \includegraphics[width=\textwidth]{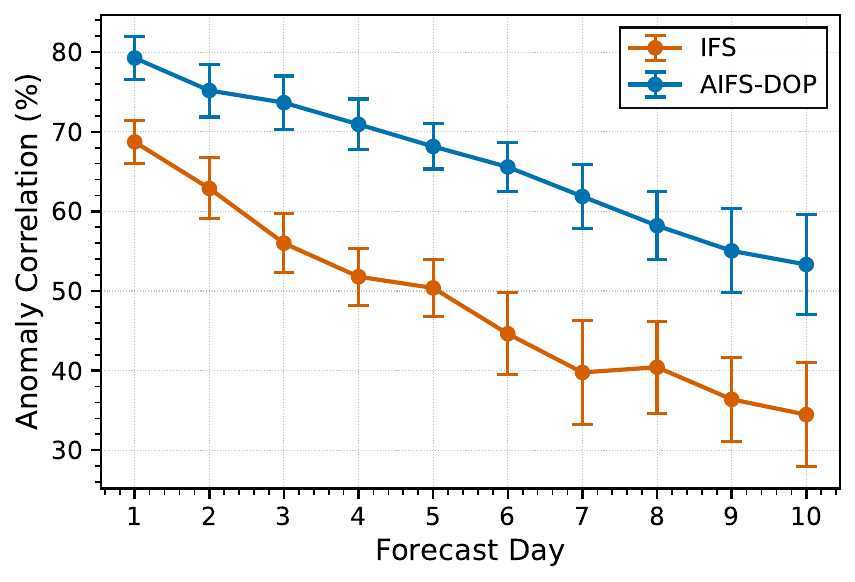}
        \caption{Tropics T-850}
        \label{fig:t850_trop_jja}
    \end{subfigure}
    \hfill 
    \begin{subfigure}[b]{0.18\textwidth}
        \centering
        \includegraphics[width=\textwidth]{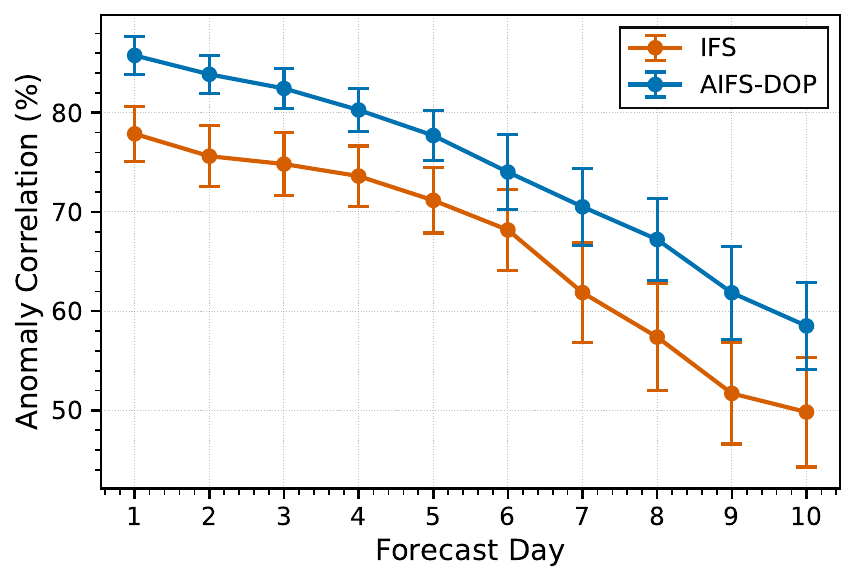}
        \caption{Tropics Z-500}
        \label{fig:z500_trop_jja}
    \end{subfigure}
    \hfill 
    \begin{subfigure}[b]{0.18\textwidth}
        \centering
        \includegraphics[width=\textwidth]{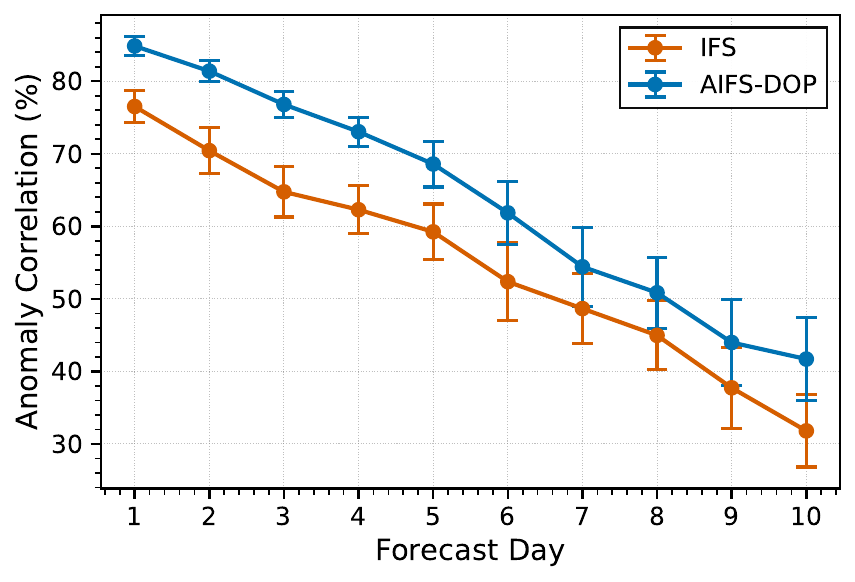}
        \caption{Tropics Wind-250}
        \label{fig:ff250_trop_jja}
    \end{subfigure}
    \hfill
    \begin{subfigure}[b]{0.18\textwidth}
        \centering
        \includegraphics[width=\textwidth]{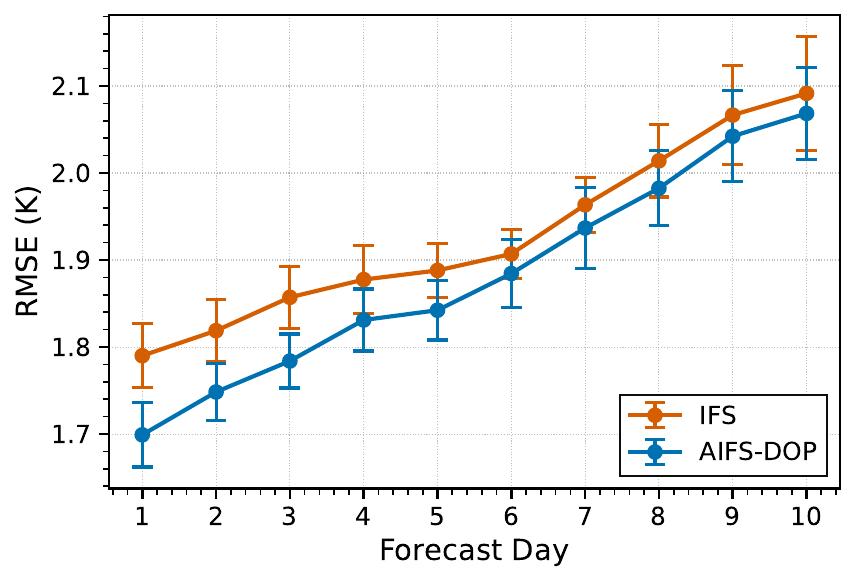}
        \caption{Tropics T2m}
        \label{fig:2t0_trop_jja}
    \end{subfigure}
    \hfill
    \begin{subfigure}[b]{0.18\textwidth}
        \centering
        \includegraphics[width=\textwidth]{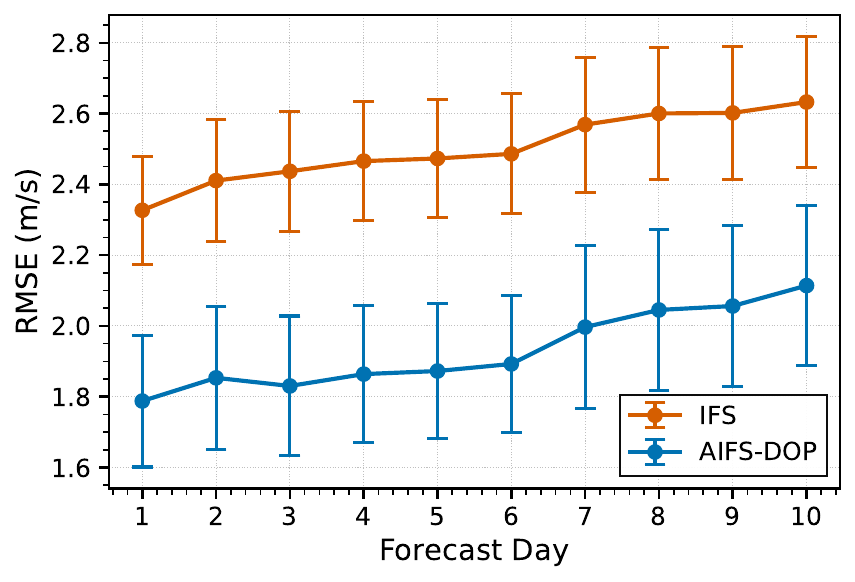}
        \caption{Tropics 10m Wind}
        \label{fig:10ff0_trop_jja}
    \end{subfigure}

    \begin{subfigure}[b]{0.18\textwidth}
        \centering
        \includegraphics[width=\textwidth]{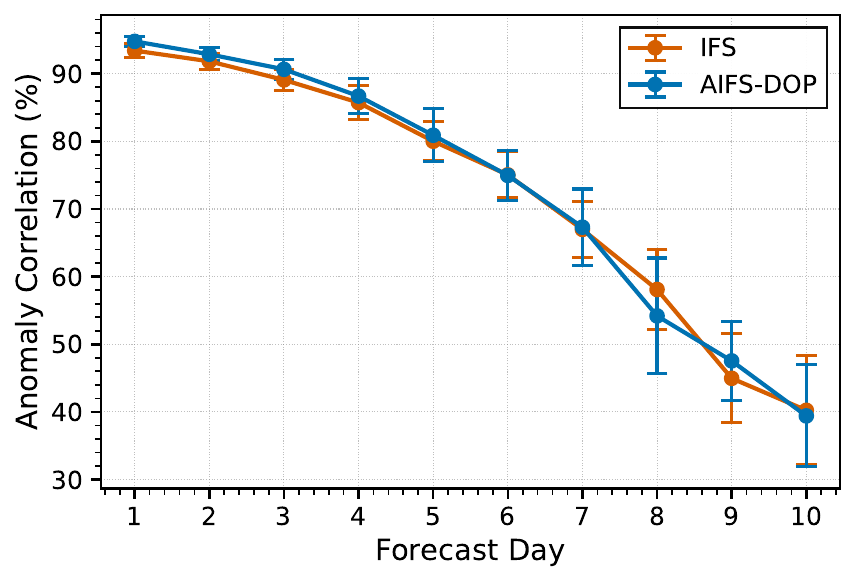}
        \caption{SH T-850}
        \label{fig:t850_sh_jja}
    \end{subfigure}
    \hfill 
    \begin{subfigure}[b]{0.18\textwidth}
        \centering
        \includegraphics[width=\textwidth]{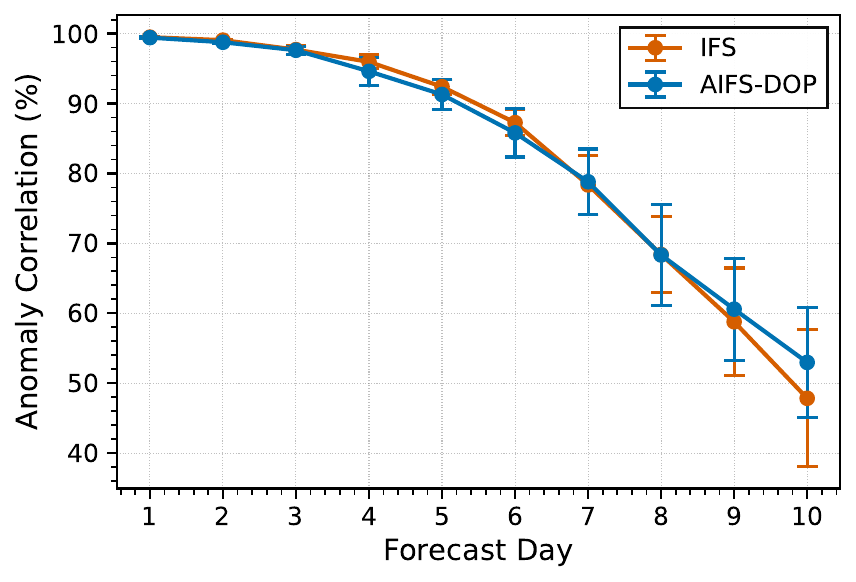}
        \caption{SH Z-500}
        \label{fig:z500_sh_jja}
    \end{subfigure}
    \hfill 
    \begin{subfigure}[b]{0.18\textwidth}
        \centering
        \includegraphics[width=\textwidth]{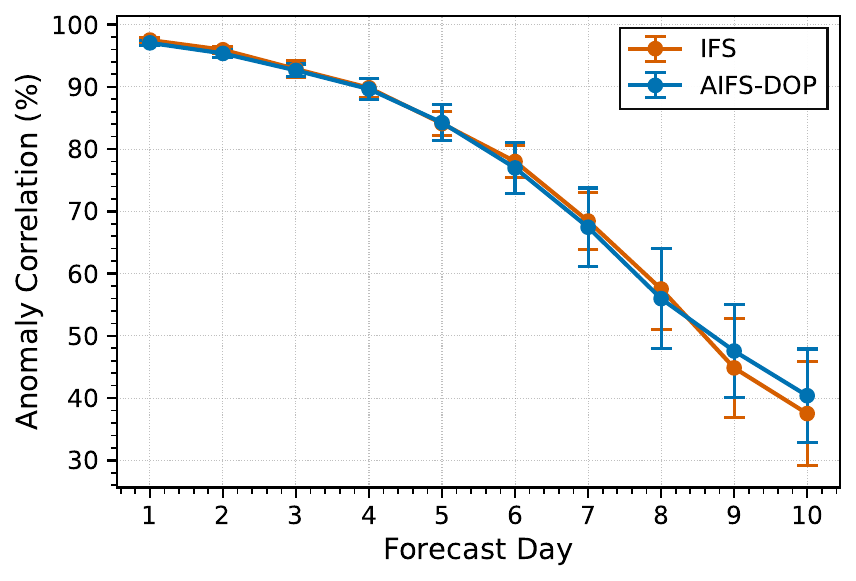}
        \caption{SH Wind-250}
        \label{fig:ff250_sh_jja}
    \end{subfigure}
    \hfill
    \begin{subfigure}[b]{0.18\textwidth}
        \centering
        \includegraphics[width=\textwidth]{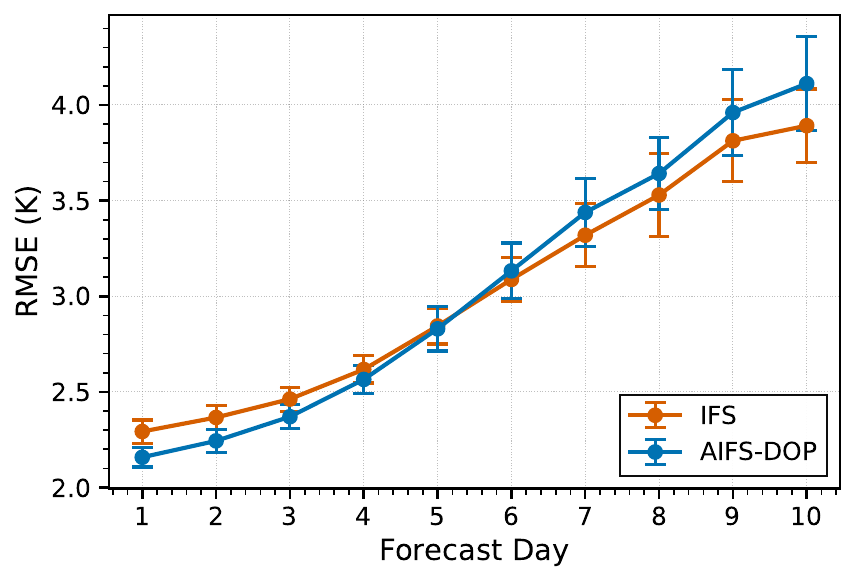}
        \caption{SH T2m}
        \label{fig:2t0_sh_jja}
    \end{subfigure}
    \hfill
    \begin{subfigure}[b]{0.18\textwidth}
        \centering
        \includegraphics[width=\textwidth]{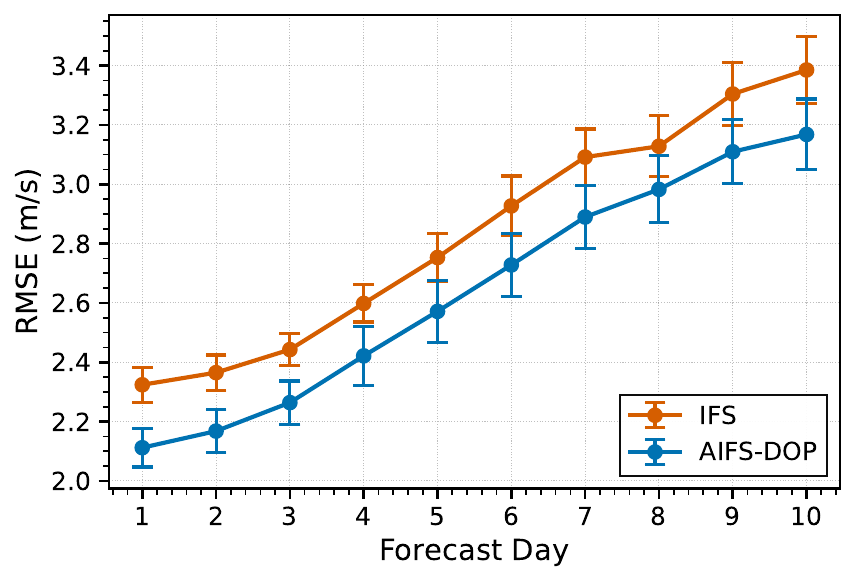}
        \caption{SH 10m Wind}
        \label{fig:10ff0_sh_jja}
    \end{subfigure}

    \caption{Upper-air anomaly correlation and surface RMSE scores computed against radiosonde and SYNOP observations respectively for June to August 2021}
    \label{fig:summer_scores}
\end{figure}

\begin{figure}
    \centering
    \begin{subfigure}[b]{0.18\textwidth}
        \centering
        \includegraphics[width=\textwidth]{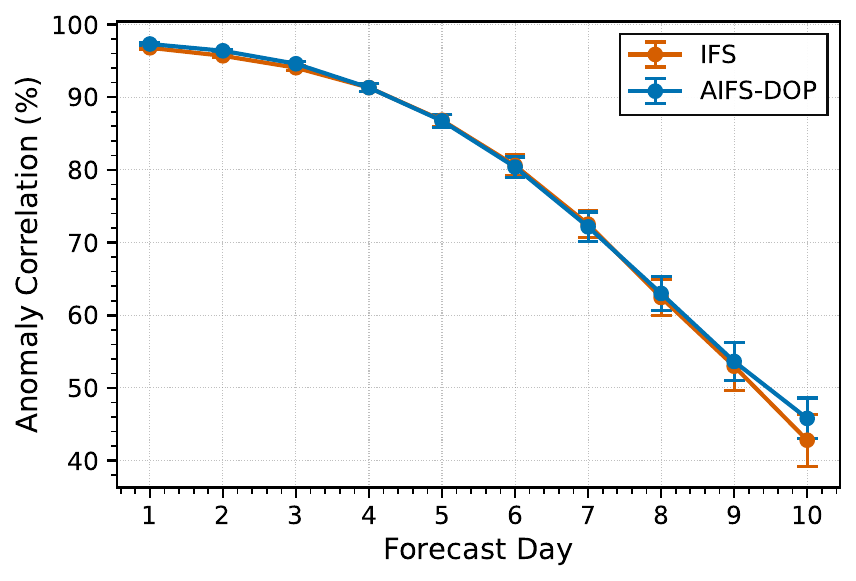}
        \caption{NH T-850}
        \label{fig:t850_nh_djf}
    \end{subfigure}
    \hfill 
    \begin{subfigure}[b]{0.18\textwidth}
        \centering
        \includegraphics[width=\textwidth]{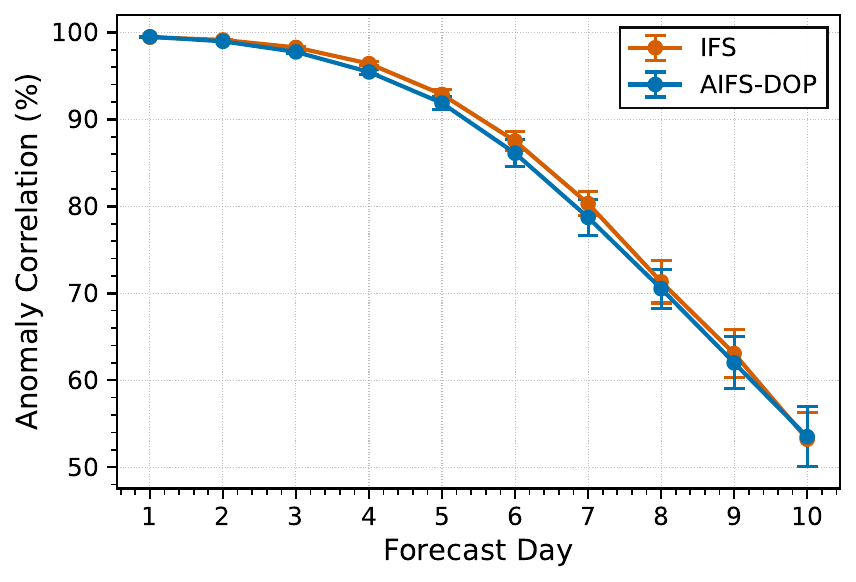}
        \caption{NH Z-500}
        \label{fig:z500_nh_djf}
    \end{subfigure}
    \hfill 
    \begin{subfigure}[b]{0.18\textwidth}
        \centering
        \includegraphics[width=\textwidth]{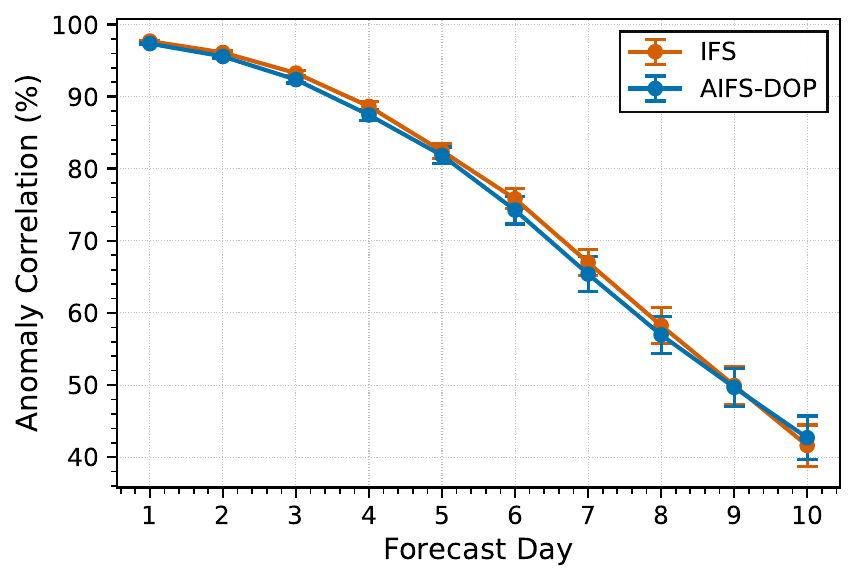}
        \caption{NH Wind-250}
        \label{fig:ff250_nh_djf}
    \end{subfigure}
    \hfill
    \begin{subfigure}[b]{0.18\textwidth}
        \centering
        \includegraphics[width=\textwidth]{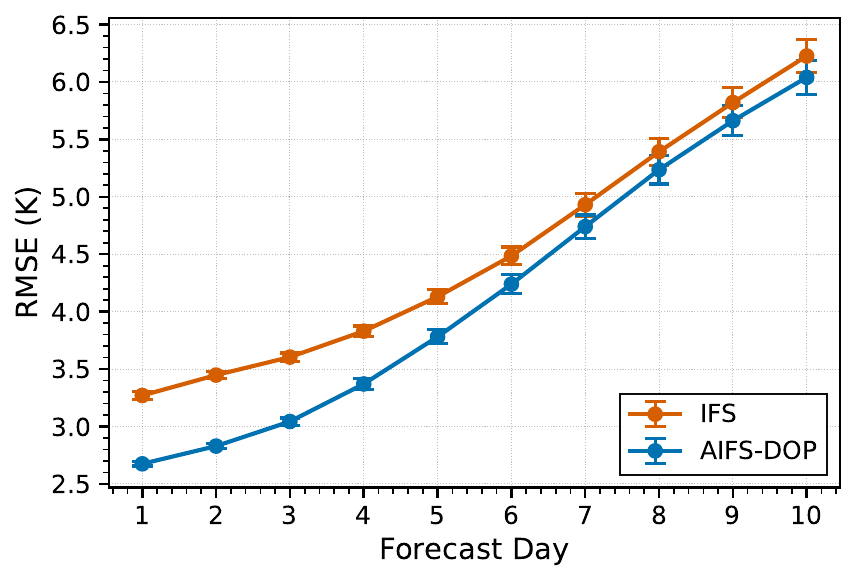}
        \caption{NH T2m}
        \label{fig:2t0_nh_djf}
    \end{subfigure}
    \hfill
    \begin{subfigure}[b]{0.18\textwidth}
        \centering
        \includegraphics[width=\textwidth]{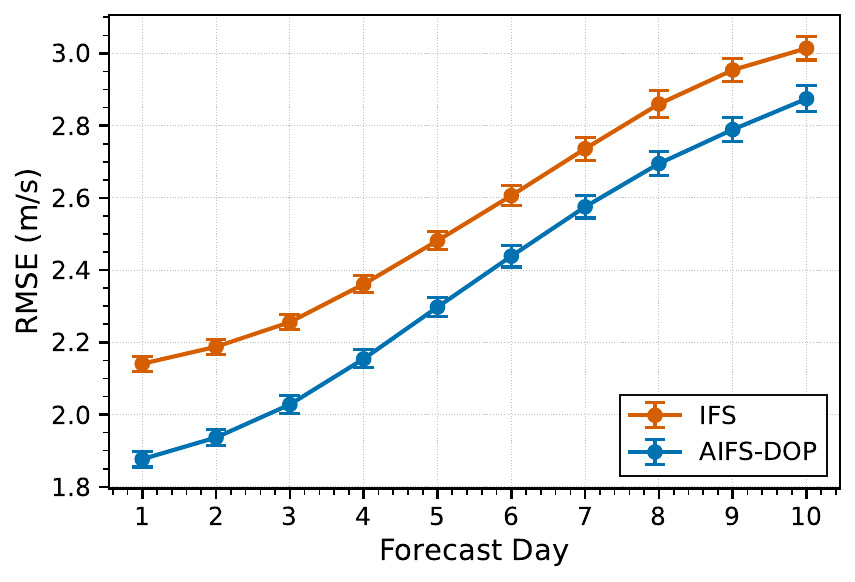}
        \caption{NH 10m Wind}
        \label{fig:10ff0_nh_djf}
    \end{subfigure}
    \begin{subfigure}[b]{0.18\textwidth}
        \centering
        \includegraphics[width=\textwidth]{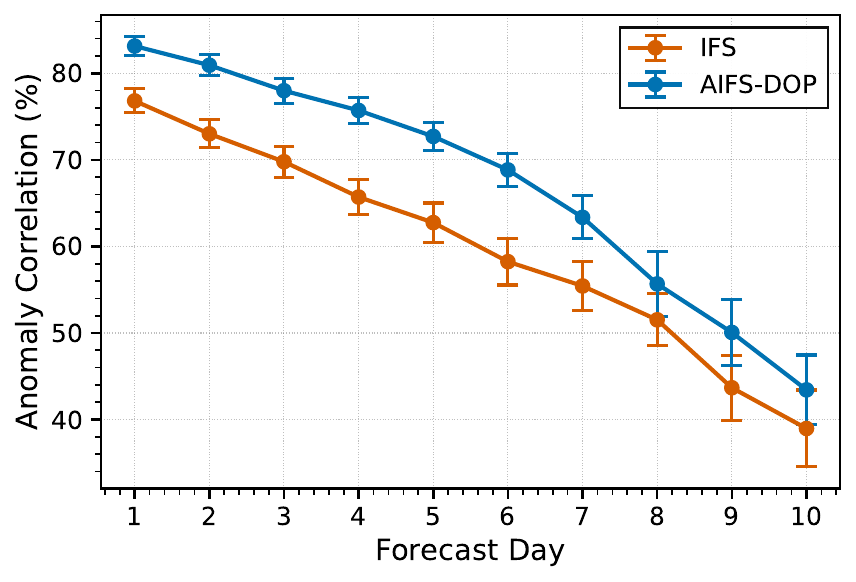}
        \caption{Tropics T-850}
        \label{fig:t850_trop_djf}
    \end{subfigure}
    \hfill 
    \begin{subfigure}[b]{0.18\textwidth}
        \centering
        \includegraphics[width=\textwidth]{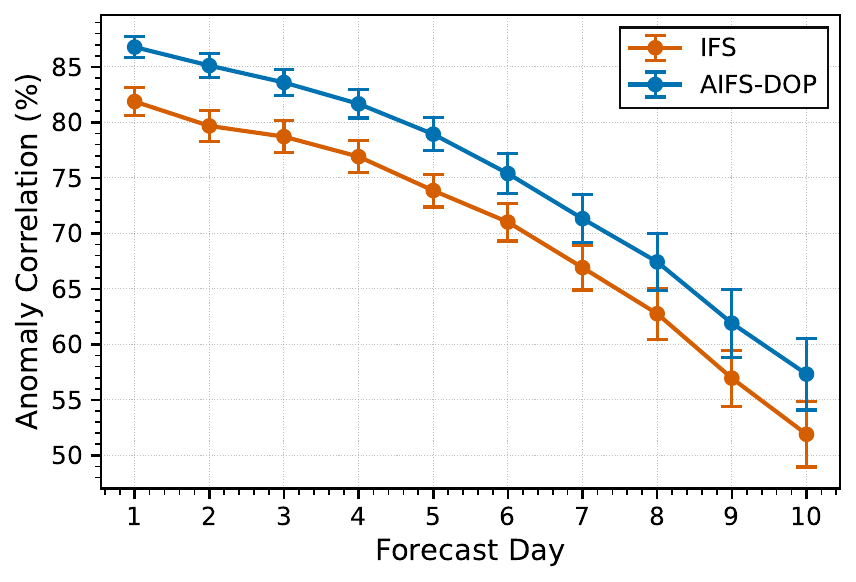}
        \caption{Tropics Z-500}
        \label{fig:z500_trop_djf}
    \end{subfigure}
    \hfill 
    \begin{subfigure}[b]{0.18\textwidth}
        \centering
        \includegraphics[width=\textwidth]{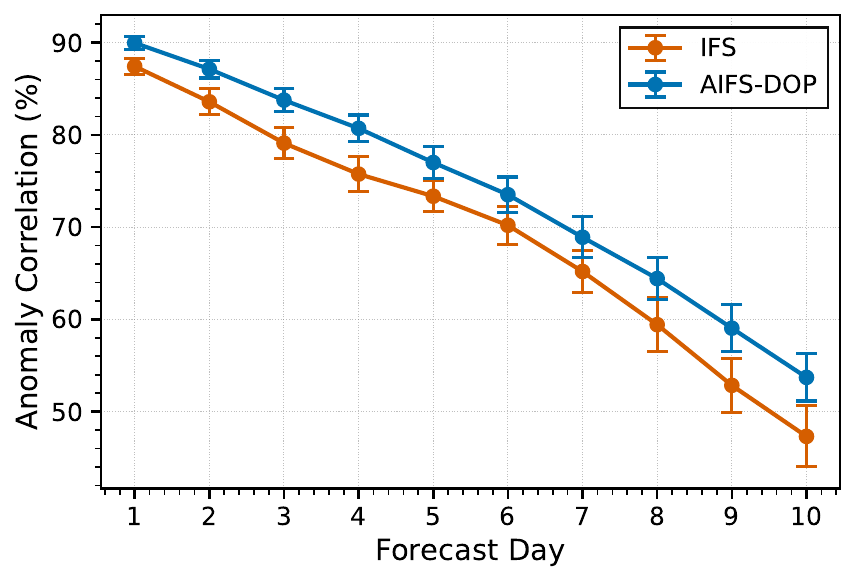}
        \caption{Tropics Wind-250}
        \label{fig:ff250_trop_djf}
    \end{subfigure}
    \hfill
    \begin{subfigure}[b]{0.18\textwidth}
        \centering
        \includegraphics[width=\textwidth]{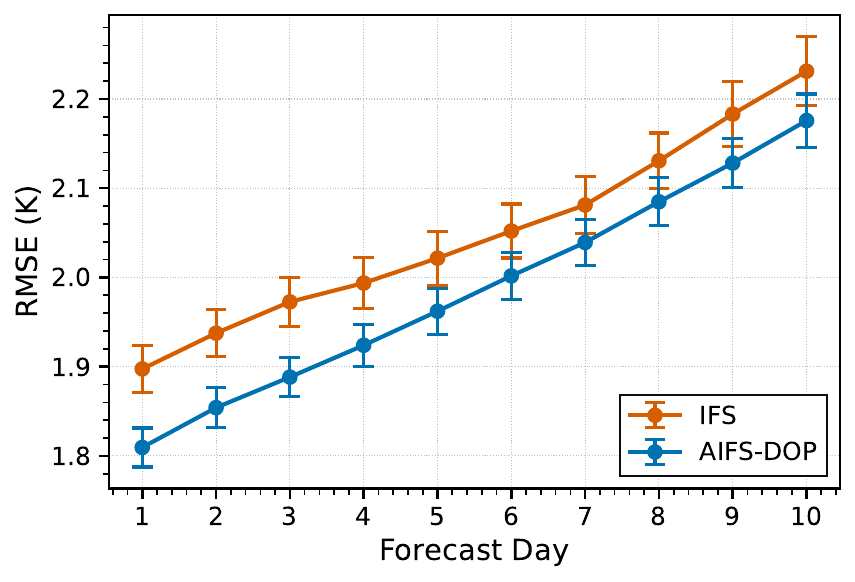}
        \caption{Tropics T2m}
        \label{fig:2t0_trop_djf}
    \end{subfigure}
    \hfill
    \begin{subfigure}[b]{0.18\textwidth}
        \centering
        \includegraphics[width=\textwidth]{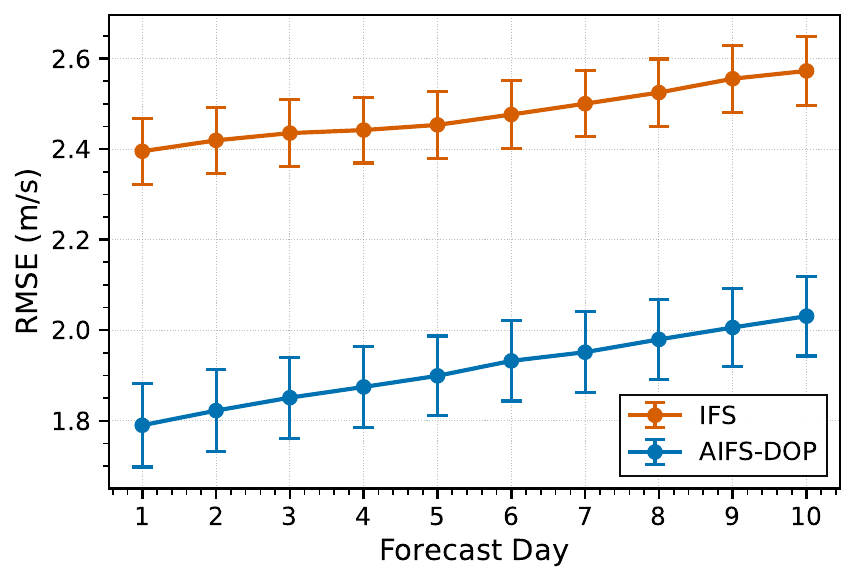}
        \caption{Tropics 10m Wind}
        \label{fig:10ff0_trop_djf}
    \end{subfigure}
    \begin{subfigure}[b]{0.18\textwidth}
        \centering
        \includegraphics[width=\textwidth]{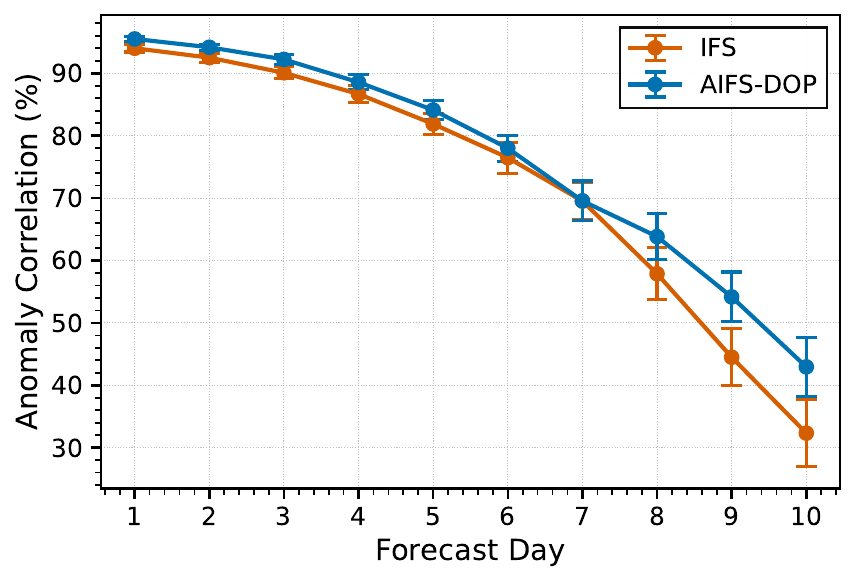}
        \caption{SH T-850}
        \label{fig:t850_sh_djf}
    \end{subfigure}
    \hfill 
    \begin{subfigure}[b]{0.18\textwidth}
        \centering
        \includegraphics[width=\textwidth]{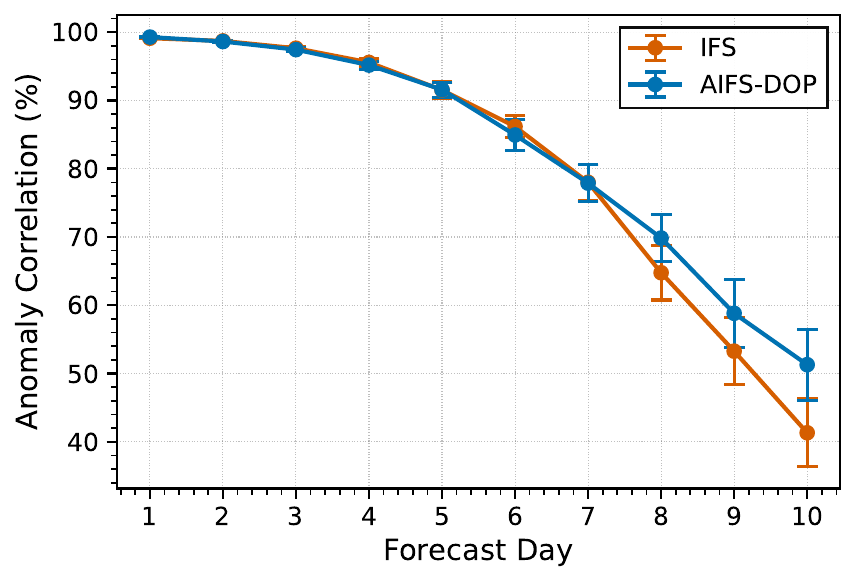}
        \caption{SH Z-500}
        \label{fig:z500_sh_djf}
    \end{subfigure}
    \hfill 
    \begin{subfigure}[b]{0.18\textwidth}
        \centering
        \includegraphics[width=\textwidth]{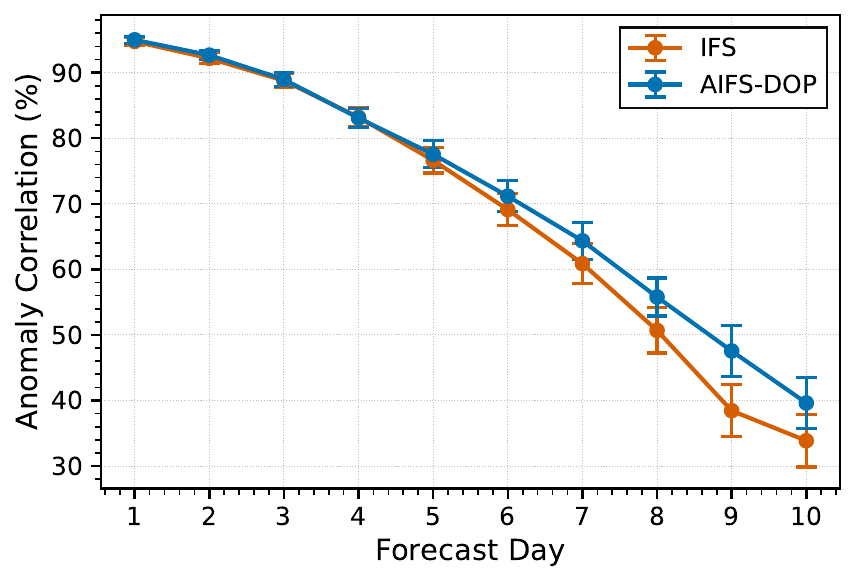}
        \caption{SH Wind-250}
        \label{fig:ff250_sh_djf}
    \end{subfigure}
    \hfill
    \begin{subfigure}[b]{0.18\textwidth}
        \centering
        \includegraphics[width=\textwidth]{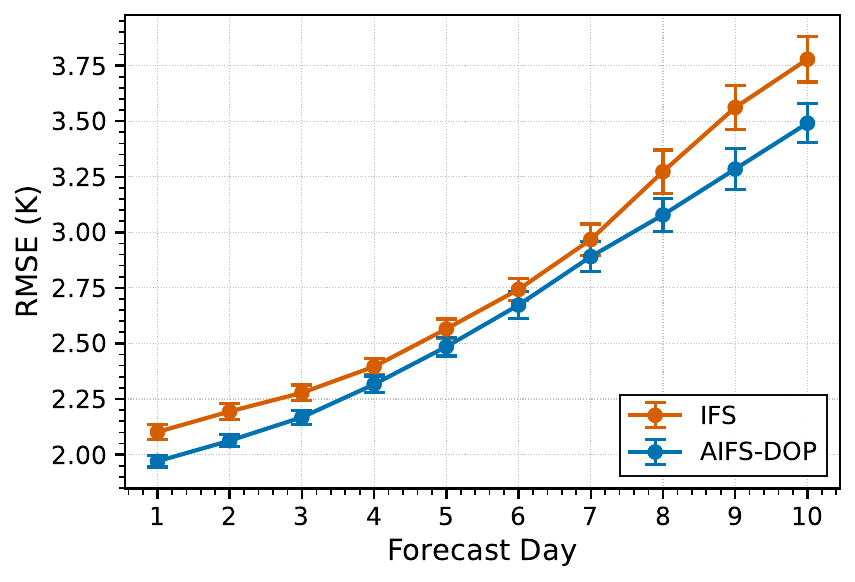}
        \caption{SH T2m}
        \label{fig:2t0_sh_djf}
    \end{subfigure}
    \hfill
    \begin{subfigure}[b]{0.18\textwidth}
        \centering
        \includegraphics[width=\textwidth]{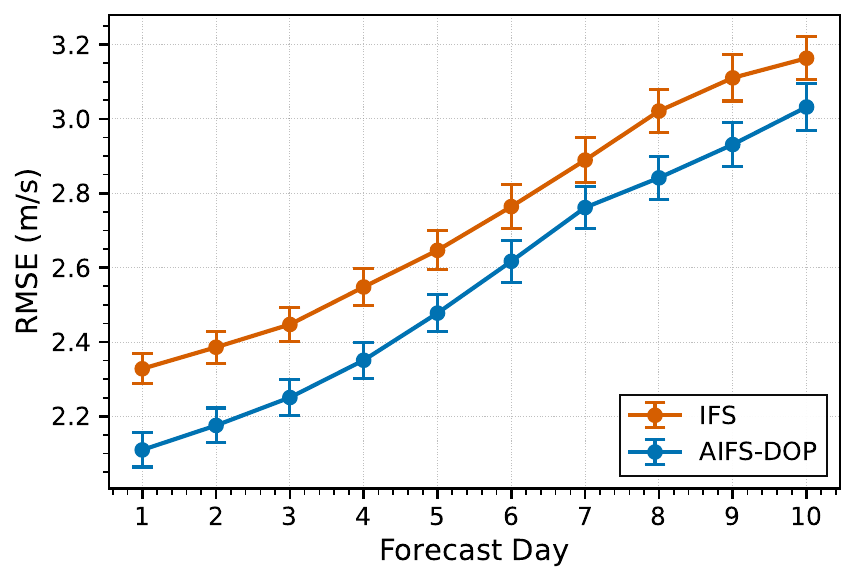}
        \caption{SH 10m Wind}
        \label{fig:10ff0_sh_djf}
    \end{subfigure}
    \caption{Upper-air anomaly correlation and surface RMSE scores computed against radiosonde and SYNOP observations respectively for December 2021 to February 2022}
    \label{fig:winter_scores}
\end{figure}

In this section we show both Northern Hemisphere Summer (JJA) and Winter (DJF) scores, in Figure~\ref{fig:summer_scores} and \ref{fig:winter_scores} respectively, to show the relative performance of AIFS-DOP in different seasons.

\end{document}